\documentclass[authoryear,12pt]{elsarticle}

\usepackage[utf8]{inputenc}     
\usepackage[T1]{fontenc}        

\usepackage{graphicx}
\usepackage{amssymb}

\usepackage{caption}
\usepackage{subcaption}

\usepackage[hidelinks]{hyperref}

\usepackage{lineno}

\usepackage{booktabs}			
\usepackage{multirow}
\usepackage{tabularx}
\usepackage{array}          
  \newcolumntype{C}{>{\centering\arraybackslash}X} 
\usepackage{hyphenat}

\usepackage[dvipsnames]{xcolor}

\biboptions{sort}

\journal{Biosystems Engineering}

\begin{document}

\begin{frontmatter}


\title{Using segment-based features of jaw movements to recognize foraging activities in grazing cattle}



\author[add1,add6]{José~O.~Chelotti}
\ead{jchelotti@sinc.unl.edu.ar}
\author[add1]{Sebastián~R.~Vanrell}
\author[add1]{Luciano~S.~Martinez-Rau}
\author[add2,add5]{Julio~R.~Galli}
\author[add3]{Santiago~A.~Utsumi}
\author[add2]{Alejandra~M.~Planisich}
\author[add2]{Suyai~A.~Almirón}
\author[add1]{Diego~H.~Milone}
\author[add1]{Leonardo~L.~Giovanini}
\author[add1,add4]{H.~Leonardo~Rufiner}

\address[add1]{Instituto de Investigación en Señales, Sistemas e Inteligencia Computacional, sinc(i), FICH-UNL/CONICET, Argentina}
\address[add2]{Facultad de Ciencias Agrarias, Universidad Nacional de Rosario, Argentina}
\address[add3]{W.K. Kellogg Biological Station and Departament of Animal Science, Michigan State University, United States}
\address[add4]{Facultad de Ingeniería, Universidad Nacional de Entre Ríos, Argentina}
\address[add5]{Instituto de Investigaciones en Ciencias Agrarias de Rosario, IICAR, UNR-CONICET, Argentina}
\address[add6]{TERRA Teaching and Research Center, University of Liège, Gembloux Agro-Bio Tech (ULiège-GxABT), 5030 Gembloux, Belgium}

\begin{abstract}
Precision livestock farming optimizes livestock production through the use of sensor information and communication technologies to support decision making, proactively and near real-time. Among available technologies to monitor foraging behavior, the acoustic method has been highly reliable and repeatable, but can be subject to further computational improvements to increase precision and specificity of recognition of foraging activities. In this study, an algorithm called Jaw Movement segment-based Foraging Activity Recognizer \mbox{(JMFAR)} is proposed. The method is based on the computation and analysis of temporal, statistical and spectral features of jaw movement sounds for detection of rumination and grazing bouts. They are called JM-segment features because they are extracted from a sound segment and expect to capture JM information of the whole segment rather than individual JMs. Two variants of the method are proposed and tested: (i)~the temporal and statistical features only \mbox{JMFAR-ns}; and (ii)~a feature selection process (\mbox{JMFAR-sel}). The \mbox{JMFAR} was tested {on signals registered in a free grazing environment}, achieving an average weighted F1-score of 93\%. 
Then, it was compared with a state-of-the-art algorithm, showing improved performance for estimation of grazing bouts (+19\%). The \mbox{JMFAR-ns} variant reduced the computational cost by 25.4\%, but achieved a slightly lower performance than the \mbox{JMFAR}. The good performance and low computational cost of \mbox{JMFAR-ns} supports the feasibility of using this algorithm variant for real-time implementation in low-cost embedded systems. The method presented within this publication is protected by a pending patent application: AR P20220100910.

\end{abstract}


\end{frontmatter}


\section{Introduction}
\label{S:1}

Precision livestock farming offers the unprecedented opportunity to improve livestock management by switching the scale of observation from herds to individuals. The individualized monitoring with regard to animal behavior could provide rich and unique information on collective events related to the health, nutrition and welfare status of animals \citep{Hodgson1998-xy}. The feasibility for deployment of animal sensing and information systems on farms \citep{Andriamandroso2016-lr} has been enhanced recently by new developments in animal-borne sensors, advanced communication and processing technologies, enhanced data analytics and visualization techniques for easy inspection of animal behavior, including behavior events and activities, location and posture \citep{Berckmans2014-qk}. However, the operational requirements for such monitoring systems, primarily device portability, robustness and low power consumption often emerges as major bottlenecks for further technological progresses \citep{Stone2020-jl}. Some of these factors depend mainly on the application and the characteristics of the proposed algorithms to solve the task. As an example, algorithms with low computational cost are preferred for continuous feeding monitoring in ruminants.

For free-grazing cattle, grazing and rumination usually represent the largest fraction of the daily time budget. A close inspection of their longitudinal changes could be useful to accurately assess animal behavior and optimize animal and pasture management. For example, decreased daily grazing and/or rumination time could be interpreted as an indicator of severe environmental stress \citep{Herskin2004-km}, anxiety \citep{Bristow2007-qe}, or emerging diseases \citep{welch1982rumination}, whereas increasing daily activity and decreasing rumination time could flag the onset of estrus \citep{Schirmann2009-ns} or calving events \citep{Schirmann2013-yq}. Similarly, increasing rumination time is associated with more saliva production and improved rumen and animal health \citep{Beauchemin1991-rp}, whereas decreasing rumination would flag a lacking effective fiber in the diets of cows \citep{beauchemin2018invited}. Grazing encloses behaviors related to feed search apprehension and consumption and is composed by three types of distinct jaw movements (JM): biting, when herbage is apprehended and severed; chewing, when herbage is comminuted; and chew-biting when a combination chewing and biting of grass occurs in a same JM \citep{Laca1992-qq, Ungar2006-ei, Galli2018-zq}. Rumination is characterized exclusively by feed chewing occurring within regular feed swallowing-regurgitation cycles. One of the distinctive characteristics of grazing and rumination is the hierarchical array of foraging activities, duration and sequence of specific JM. Whereas chewing and biting in a single JM last about a second or a fraction of a second, the sequence of JM that defines a grazing or rumination bout usually expands from several minutes to hours \citep{Andriamandroso2016-lr}.

There are several remote sensing methods to monitor foraging behavior. Among most popular methods are the accelerometers \citep{pavlovic2022behavioural, pavlovic2021classification, shen2020automatic, benaissa2019use}, inertial measurement units \citep{li2022classification, sakai2019behavior, carslake2020machine, peng2019classification} and nose-band pressure sensors \citep{chen2022recognition, zehner2017system,benaissa2019classification,eslamizad2018development}. 
These acquisition methods 
provide straightforward information of whole body movements or movement of specific body parts, such as leg, neck or head movements \citep{lee2021wearable}. The recognition of animal behaviors could be improved using various sensors in different locations of the animal \citep{benaissa2019use,brennan2021classifying}.
This approach is not always practical since it is more laborious for both humans and cows and more expensive \citep{riaboff2020development}.
A commercial system formed by a nose-band pressure sensor with an accelerometer in a halter measures foraging behaviors and provide detailed information associated.
The raw data collected has to be processed by a software in a computer to obtain behaviors information, thus demanding handlers \citep{werner2018evaluation, steinmetz2020validation}.
Accelerometers and inertial measurement units are used to recognize feeding, posture, and locomotion behaviors of grazing cattle.
In last decade, many methods to recognize the two main foraging activities (grazing and rumination) using machine learning techniques have appeared \citep{dutta2015dynamic,bishop2014investigation}.
They mainly differ in the behaviors of interest, sensor location and model engineering \citep{riaboff2022predicting}.
In this sense, the pattern recognition methods proposed by \citet{rahman2016comparison} and \citet{Smith2016-xb} distinguished both foraging behaviors and others locomotion and posture behaviors.
Moreover different accelerometer locations were evaluated to recognize grazing and rumination by \citet{rahman2018cattle}.
The work of \citet{riaboff2020development} discriminated foraging behaviors in particularly posture positions.
They were able to distinguish between standing and lying rumination among many other behaviors.
\citet{li2021data} expand the monitoring of feeding and others behaviors from free ranging environment to indoor environment employing a convolutional neural network.
This deep learning approach performed good in both environment but it was computational expensive for an implementation in an embedded device.
In fact, a low computational cost pattern recognition system using a small set of features and classical machine learning classifier was proposed by \citet{arablouei2021situ}.
Previous works reached good performance, but a deeper analysis of model generalization is still required \citep{riaboff2022predicting}.
Additionally, accelerometer based sensors are heavily dependable on the correct location, orientation, and fastened by making it not easily reproducible \citep{li2021data,kamminga2018robust}.
Thus, head-placed microphones are usually preferred over accelerometers, inertial measurement units, and nose-band pressure sensors. The good robustness to free ranging environmental noises allows the precise monitoring of foraging behavior \citep{martinez2022robust}.


The acoustic method has obtained very good results in recognizing and classifying JM and cattle grazing foraging activities. \citet{Vanrell2018-zj} proposed an off-line acoustic method called regularity-based acoustic foraging activity recognizer \mbox{(RAFAR)} that relied on autocorrelation for recognition of grazing and rumination bouts. The main disadvantage of this algorithm is the requirement of the entire sound record to achieve satisfactory performance results. More recently, \citet{chelotti2020online} introduced \mbox{BUFAR}, an alternative algorithm to \mbox{RAFAR} to enable on-line processing of sound signals. Both, \mbox{RAFAR} and \mbox{BUFAR} achieved better performance compared to commercial rumination collars \citep{Vanrell2018-zj, chelotti2020online}, yet the two have also shown room of improvement for the recognition of grazing and rumination.

The main objective of this work is to develop and test a new on-line method for improved classification of grazing and rumination bouts in free grazing cattle. The new method will use broad sets of temporal, statistical and spectral features of the sound signal and it is therefore expected to minimize previous grazing and rumination misclassifications, without significantly incurring in additional computational costs.
The new features are called JM-segment features because they are extracted from a sound segment to capture JM insight of the whole segment rather than individual JMs.
The main benefits for the new method are: (i)~the grazing and rumination classification is independent of the JM~classification, thus avoiding the errors associated with this stage \citep{chelotti2018pattern}; and (ii)~the ability to train an on-line model
without the need of using previously labeled JM, which is a significant improvement that enable to train the method on large recording databases. The paper is organized as follows: Section~2 introduces the proposed algorithm and the new set of features.
Databases used for model testing, the variants of the proposed method and the measures of performance to assess model suitability are described in Section~3. Section~4, describes the main study results with a focused discussion presented in Section~5. Finally, a brief conclusion is presented in Section~6.

\section{Material and methods}
\subsection{Databases}
In this study, two independent databases composed of sound recordings of dairy cows were used.
The first database (referred as DbM)
was generated from measurements obtained from July 31 to August 18, 2014, at the Michigan State University's Pasture Dairy Research Center facility at the W.K. Kellogg Biological Station, located in Hickory Corners, Michigan.
Sound tracks were recorded on five experimental, high-production multiparous Holstein cows (652.0 $\pm$ 5.4 kg of live weight; 2.3 $\pm$ 0.5 calvings; 102.0 $\pm$ 14.6 days in milk; 41.2 $\pm$ 3.9 kg of daily milk production) managed as part of a larger herd ($\sim$140 cows). The fieldwork was carried out following protocols for robotic milking routines (3.0 $\pm$ 1.0 daily milkings), using two Lely A3-Robotic milking machines (Lely Industries NV, Maassluis, The Netherlands). The cows had permanent access to pastures with a predominance of perennial ryegrass (Lolium perenne) and white clover (Trifolium repens), or with a predominance mixture of orchardgrass (Dactylis glomerata), tall fescue (Lolium arundinacea) and white clover. The pastures were grazed in two shifts, from 10:00 to 22:00 and from 22:00 to 10:00. In each shift cows freely accessed a new paddock with the similar level of daily allowance (15 kg of dry matter/cow). Consequently, the average daily pasture allowance was 30 kg of dry matter/cow. The forage quality of hand plucking samples was determined through five parameters: in vitro of dry matter digestibility (IVDMD, Daisy II, Ankom Technology Corp.) the average content of crude protein (CP, 4010 CN Combustion, Costech Analytical Technologies Inc., Valencia, CA), Neutral Detergent Fiber (NDF), Acid Detergent Fiber (ADF, Fiber Analyzer, Ankom Technology Corp., Fairport, NY), and Acid Detergent Lignin (ADL). The values (mean $\pm$ standard deviation) estimated for IVDMD, CP, NDF, ADF and ADL were 78.1 $\pm$ 3.0\%, 18.7 $\pm$ 2.5\%, 49.3 $\pm$ 4.5\%, 25.7 $\pm$ 2.0\%, 3.3 $\pm$ 0.8\% respectively.

Twenty four hour continuous acoustic recordings obtained on 6 non-consecutive days were used. In each of the days the foraging behavior of the five cows was recorded. Five independent equipment devices (halter, microphone and recorder) were used, and they were rotated on the five cows throughout the six experimental days. Each equipment included a directional microphone pressed to the forehead of the animal connected to a digital recorder. The microphones were placed against the forehead of the cow, held and protected by an elastic band, to prevent microphone movement, isolate noise caused by wind and protect microphones from sources of friction and scratches. All recordings were performed with a 44.1~kHz sample rate and 16-bit resolution, providing a nominal recording bandwidth of 22~kHz and a dynamic range of 96~dB. The records were saved in Waveform Audio (WAV) file format.

The audio recordings belonging to the second database (referred as DbZ) were obtained in October 2014 during grazing feeding trials conducted at the dairy farm of the Campo Experimental J. Villarino of Universidad Nacional de Rosario, Argentina. The protocols for handle and care of the cows were previously evaluated and approved by the Committee on Ethical Use of Animals for Research of the Universidad Nacional de Rosario. 
Five lactating multiparous 3–5~year-old Holstein cows weighing $570\pm40$ were managed along with a larger dairy herd ($\sim$150 cows).
The foraging behaviour of the experimental cows grazing on alfalfa (\textit{Medicago sativa}), fescue (\textit{Festuca arundinacea}) and prairie grass (\textit{Bromus catharticus}) mixed pastures, were continuously monitored for 24~h during six non-consecutive days.
Each cows used one of the five equipment devices previously used to generate the DbM database.
The equipment devices were randomly assigned to the cows at the beginning of the experiment and rotated daily according to a 5~x~5 latin-square design.
Particularly at outside of the barn, the nature and environmental conditions during this experiment have generated more noisy audio signal than those obtained in the DbM database, associated to machines, vehicles, winds and, subjection and movements of the equipment devices, among others \citep{martinez2022robust}.

Two experts with long experience in foraging behavior detection and digital analysis of acoustic signals identified the activities through direct observation and listening of recordings, using the Audacity\footnote{https://www.audacityteam.org/} program. During daylight hours, the activities (grazing, rumination, resting, drinking, etc.) of each of the cows were checked out every 60 minutes. This information was used to confirm or correct the grazing and rumination times obtained from Audacity sound records. Activity blocks were identified, classified and labeled, either as grazing, rumination, or other activities.

\label{S:2}
\subsection{Background: Current acoustic method}

In a previous work, \citet{chelotti2018pattern} introduced a JM~recognition algorithm that estimates sequences of JM using the sounds produced by ruminants during foraging activities. It contains three distinct elements: (i)~a signal conditioning stage, which attenuates the effects of environmental noises and disturbances, (ii)~a set of extracted acoustical features and (iii)~a machine learning model that provides the algorithm with excellent discrimination capabilities. Based on this algorithm, a foraging activity recognizer called \mbox{BUFAR} was proposed by \citet{chelotti2020online}. It analyses groups of JM to recognize grazing and rumination bouts. It is comprised of (i)~a buffer segment, which comprises a sequence of estimated JM along a fixed period of time, (ii)~a set of descriptive statistical features for JM in the buffer that includes JM~rate and proportions for JM~types, and (iii)~a machine learning model to enable discrimination capabilities of the algorithm. Accordingly, rumination bouts were identified and determined by regular cycles of thirty to sixty pure chews, occurring at a near constant rate of one JM per second, and followed by brief sound pauses for feed cud swallowing and regurgitation \citep{galli2020discriminative}. On the other hand, grazing bouts were determined by uneven mixtures of bites, chews and chew-bites sequences that had no particular order and occurred at near constant rate of one JM per second \citep{chelotti2020online}.

\subsection{Recognition system}
\begin{figure}[t]
       \centering
       \includegraphics[width=\textwidth]{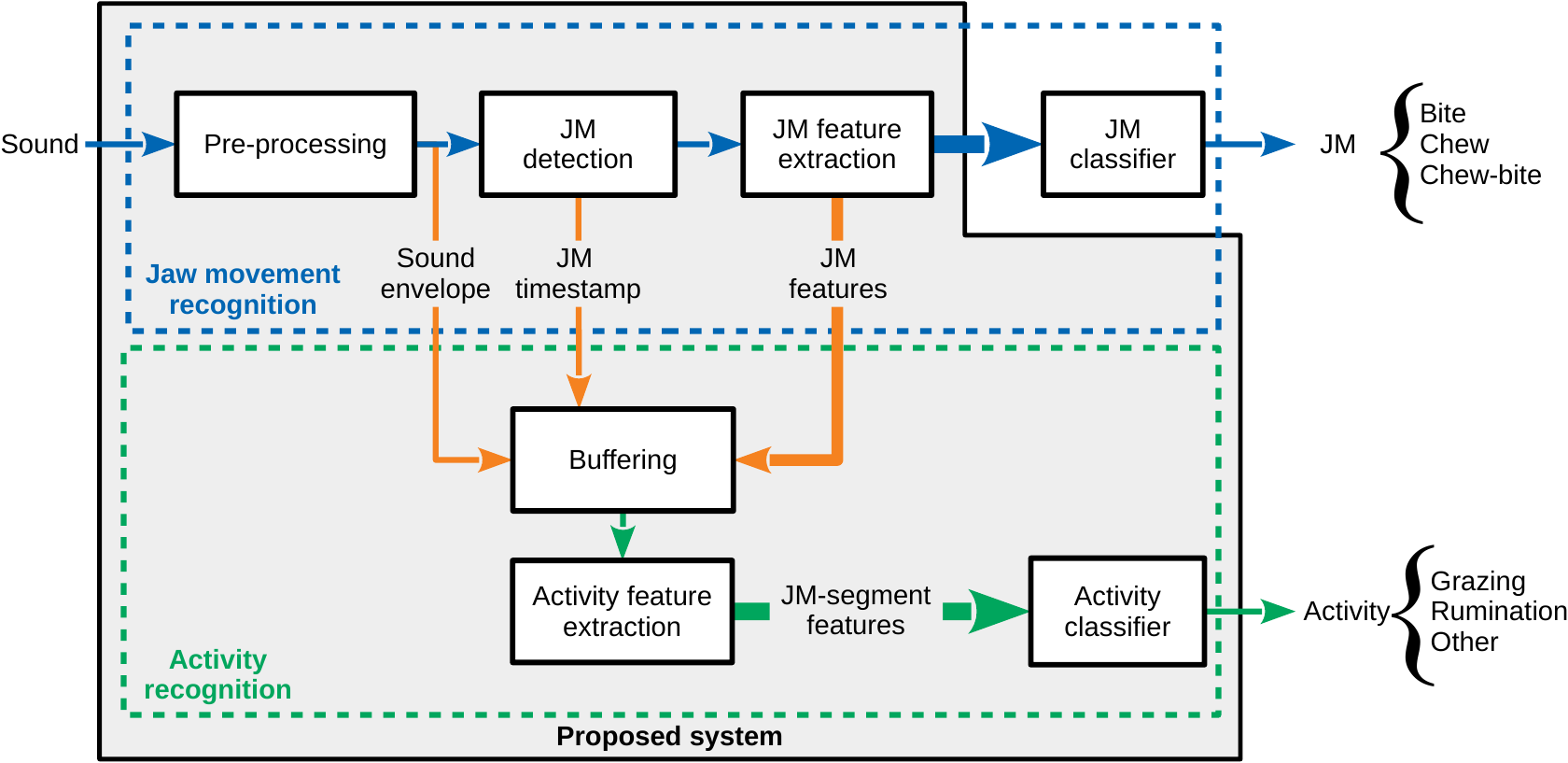}
       \caption{
       General diagram of the architecture. Blocks inside the gray box are part of the proposed system (JMFAR).
       The thickness of the arrows indicates the size of the feature vector.}
       \label{fig:blockDiagram}
\end{figure}

A block diagram of the proposed system is shown in Figure~\ref{fig:blockDiagram}. It consists of a JM~recognizer \citep{chelotti2018pattern} and an activity recognizer, which is inspired by \mbox{BUFAR} architecture.
To improve the performance of the activity recognizer, a novel set of discriminative features were developed for the activity classifier.
The system analyses the acoustic signal in non-overlapping segments of 5~min length to determine the type of activity performed in the segment. 
The use of non-overlapped segments has a computational cost lower than the strategy of overlapped segments, and we observed similar performance for both cases.
First, in a pre-processing stage the acoustic signal is conditioned to remove noises and disturbances, thereby improving the signal-to-noise ratio. For this stage, a least mean
square (LMS) filter was used \citep{widrow1975adaptive}. This adaptive filter has
proven to be useful for removing trends at low computational cost.
Then, a decimation process is applied and the sound envelope is obtained. After this, working with a low-frequency signal (sound envelope) will reduce the computational cost of the algorithm in the following stages.
In the next stage, the candidate JM is detected (JM~detection stage) in order to divide the sound envelope, isolate the JM, and extract the JM~features. The signal envelope, the timestamp of the detected JM and the JM~features are stored in the segment buffer and then processed to extract the novel set of activity features. These features are analyzed by the activity classifier to determine the activity performed by the animal within the time segment.
The new features are called JM-segment features because they correspond to a sound segment and capture JM insight of the whole segment rather than individual JMs features.
Their computation only requires the detection of events, regardless of their type (i.e. bite, chew or chew-bite).
Unlike the BUFAR method, this approach has the advantage of being insensitive to JM~classification errors.
Although the JM~type is not used for the system, it might be obtained as an extra output from the JM~classifier.
Both classifiers (JM and activity) are implemented using a multilayer perceptron \citep{Bishop2006-ui}. For both classifiers, the optimized hyperparameters were the learning rate and the number of neurons in the hidden layer. For the activity classifier, the MLP structure consisted of 21 neurons in the input layer, 20 in the hidden layer and 3 in the output layer.

\subsection{Proposed activity features}

In order to improve recognition ability, we propose a set of temporal, statistical and spectral features of JM for the foraging activity recognition, which are named JM-segment features.
Previous JM~features included in \mbox{BUFAR} were the proportion of chews (f1)\footnote{Numbers in brackets indicate the ordinal number of the feature within the set.}, the proportion of bites (f2), the proportion of chew-bites (f3) and JM~rate (f4), which may contain relatively low information and discriminatory power. The new set of JM-segment features is composed by:

\begin{itemize}
    \item {JM~rate : is the proportion of detected events (f4) within a given sound signal segment.}
    \item {Statistics of JM~amplitudes : includes mean (f5), standard deviation (f6), skewness (f7) and kurtosis (f8) for maximum amplitudes of JM that occur within a given sound signal segment.}
    \item {Statistics of JM~durations : including mean (f9), standard deviation (f10), skewness (f11) and kurtosis (f12) for duration of JM that occur within a given sound signal segment.}
    \item {Spectral feature of the sound envelope : defined as energy in the 1.0-to-1.5~Hz band (f13) relative to the total energy of a given sound signal segment.}
    \item {JM~jitter variants : variation of the period between two consecutive peaks (in this case JM) within a given signal segment \citep{farrus2007jitter}. Alternative variants for this parameter include: absolute jitter (f14), relative jitter (f15), five-point period perturbation quotient (PPQ5) jitter (f16), and standard deviation of jitter (f17).}
    \item {JM~shimmer variants : variation of maximum amplitude between two adjacent peaks (in this case JM) within a given signal segment \citep{farrus2007jitter}. Alternative variants for this parameter include: absolute shimmer (f18), relative shimmer (f19), five-point amplitude perturbation quotient (APQ5) shimmer (f20) and standard deviation of shimmer (f21).}
    \item {Temporal features of the JM~tachogram :
    the tachogram is the representation, in the form of a discrete sequence, of the temporal distance between consecutive peaks.
    Thus, patterns of regular JM~sequences are used for activity classification. This feature indicates the rate for JM~intervals with lengths between 3 and 10 s (f22).}
    \item {Spectral features of the JM~tachogram : is the energy in the 0.017-to-0.020~Hz (f23) and 0.00-to-0.02~Hz (f24) bands relative to the total energy of the tachogram signal and provides information about the swallowing-regurgitation events.}
\end{itemize}

\begin{figure}
       \centering
       \includegraphics[width=.5\textwidth]	{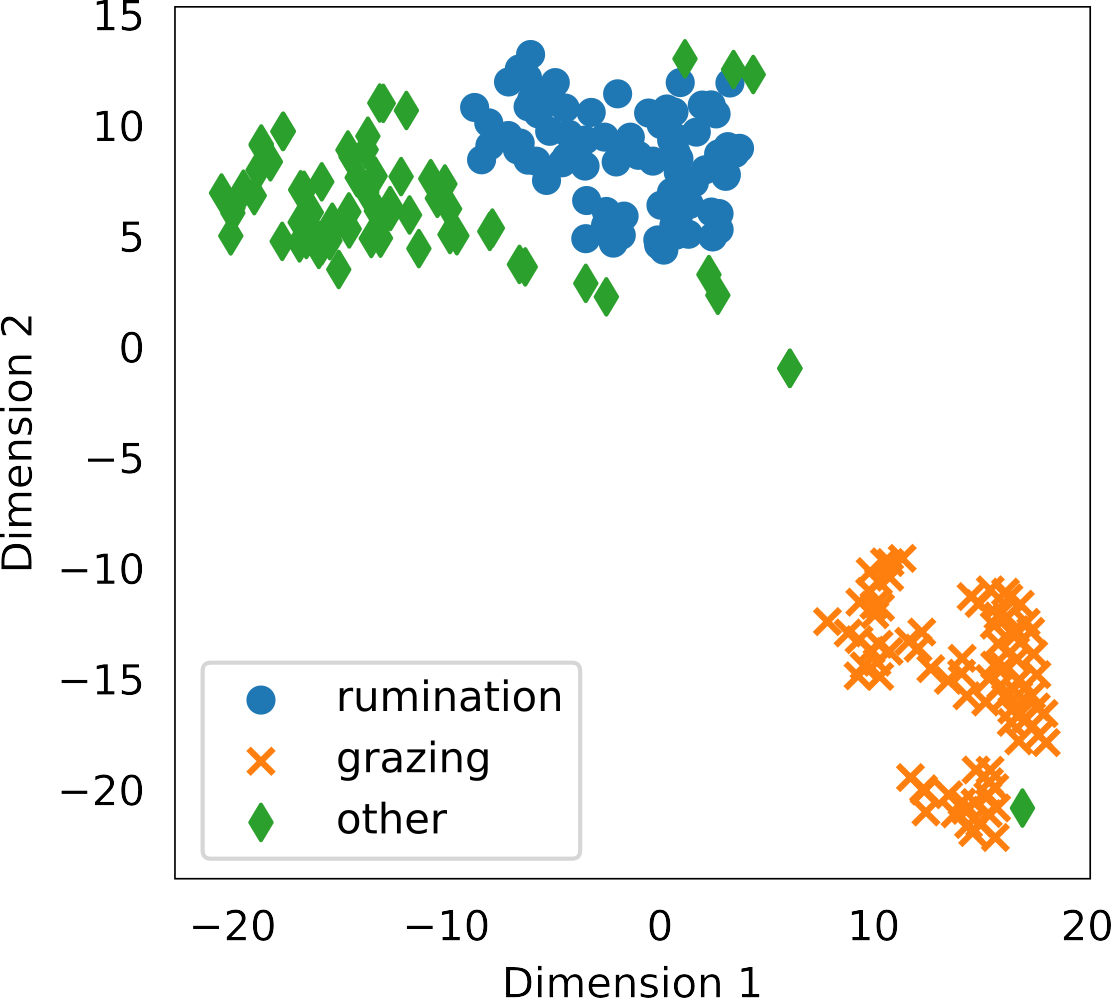}
       \caption{t-SNE analysis of JM-segment features.}
       \label{fig:tsne}
\end{figure}

A qualitative analysis of the effect of proposed features on class separation was performed. The data from the training set was used in the figures of this section. In total, four free grazing sessions of four different animals were considered, totaling about 24 hs of recordings.
Figure~\ref{fig:tsne} shows the result of a dimensionality reduction analysis \citep{van2012visualizing} applied to the set of JM-segment features proposed in this work. Grazing and rumination are grouped in clear clusters without mixed data points between them. A few proportions of activities other than grazing or rumination are minimally mixed with rumination data points.

\subsubsection{Statistical features for amplitude and duration}

\begin{figure}
       \centering
       \includegraphics[width=.7\textwidth]	{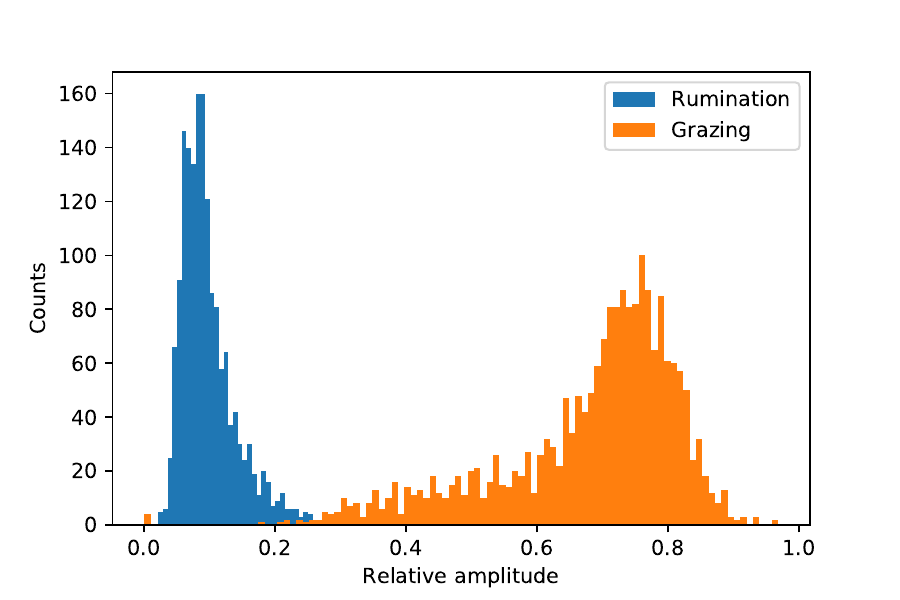}
       \caption{Histogram showing the amplitude distribution of JM associated with grazing and rumination.}
       \label{fig:histograms}
\end{figure}

The foraging behavior of ruminants can be characterized by the temporal sequence of JM, with sounds that vary according to given animal factors, foraging events and biochemical and physical properties of the feed eaten. For example, bites are characterized by a loud sound with a high amplitude peak, when forage is apprehended between incisors and the dental pad, and severed. Grazing chews produce loud sounds with lower intensity than bites. Composite chew-bites combine sound characteristics of grazing chews (first) and bites (later) during the same JM. Additionally, whereas exclusive bites and chews have a similar short duration, chew-bites have longer sound signals \citep{chelotti2018pattern}. 
Since grazing is composed by a mixed sequence of bites, chews and chew-bites, the amplitude frequency distribution peaks and spreads over a broader high amplitude region (Figure~\ref{fig:histograms}). Conversely, ruminating chews are characterized by very low intensity sounds, associated probably to the wetness, fragmentation and partial degradation of feed being processed (rumen turnover is at least fifteen hours). Since rumination is composed of exclusive rumination chews, the amplitude frequency of rumination peaks in the very low amplitude region and is narrowly distributed around peak frequency (Figure~\ref{fig:histograms}). Due to the presence of chew-bites, the event duration distribution of grazing has a peak on the long duration region and it is distributed over a wide range of values. The duration histogram of rumination has similar characteristics to its amplitude histogram. Considering the asymmetric nature of amplitude and duration histograms, several moments (mean, standard deviation, skewness and kurtosis) were evaluated and considered as features.

\subsubsection{Spectrum of the sound envelope}

The spectrum of the sound envelope helps to discriminate among, grazing, rumination and other activities. This spectrum characterizes the temporal behavior of events according to the periodicity and variability of the type of activity an animal performs. The spectra of rumination sound segments show a very narrow frequency band determined by the low periodicity and low amplitude of rumination chew. On the other hand, the spectra of grazing sound segments show a broader frequency band determined by the irregular and mixed nature and higher amplitude of chews, bites and composite chew-bites during grazing.

\begin{figure}
       \centering
       \includegraphics[width=.7\textwidth]	{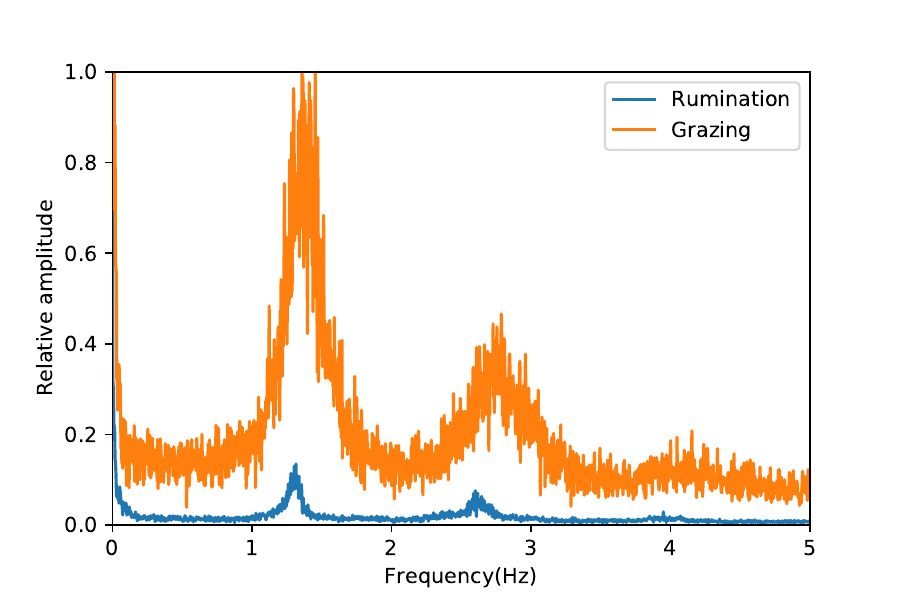}
       \caption{Spectra of grazing and rumination sound segments.}
       \label{fig:spectSignals}
\end{figure}

Figure~\ref{fig:spectSignals} shows the average spectra of typical sound segments corresponding to grazing (orange) and rumination (blue). The peaks of relative amplitude correspond to the fundamental frequency (and its harmonics) of JM, which could be useful to differentiate among grazing, rumination, and other activities. Further, the ratio between the energy in the band 1.0-to-1.5~Hz and the total sound segment energy was proposed as an additional discrimination feature. In this context, activities other than grazing and rumination are not expected to present this characteristic spectral pattern, as such activities may include periods of silence or different types of environmental noise.

\subsubsection{Tachogram}

\begin{figure}
       \centering
       \includegraphics[width=.8\textwidth]	{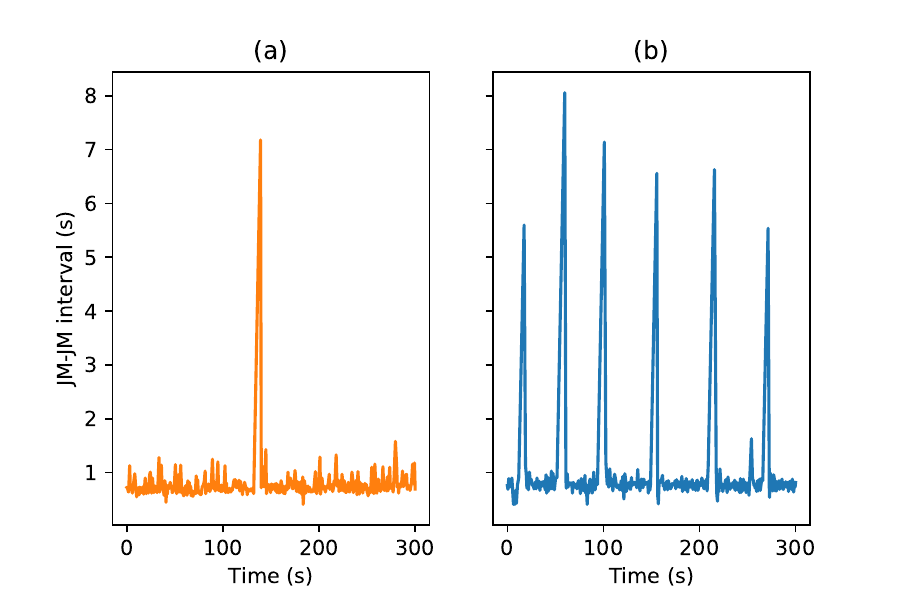}
       \caption{Tachograms of 
        five minutes of (a) grazing and (b) rumination. The JM-JM interval refers to the time between two consecutive jaw movements.}
       \label{fig:tachograms}
\end{figure}

A ruminant typically performs a bundle of JM per minute (between thirty to sixty JM) and the rate variability of JM depends on the activity being performed. The timestamps of detected events can be arranged into tachograms showing the inter-event periodicity time-series (Figure~\ref{fig:tachograms}). Grazing is composed by a sequence of regular JMs and less frequent and irregularly distributed long pauses associated with walking or food searching (Figure~\ref{fig:tachograms}.a). Thus, the rate variability of JM for grazing shows a great frequency of inter-events periods around one second (similar inter-events between successive JMs) with occasional peaks of longer duration associated with exclusive food searching \citep{bailey1996mechanisms}. On the other hand, rumination is composed by groups of regular chews followed by periodic pauses of several seconds for feed bolus swallowing and regurgitation. 
Thus, the rumination tachogram will show values around one second (the period between successive JMs) and regular peaks of three or more seconds corresponding to swallowing and regurgitation pauses
(Figure~\ref{fig:tachograms}.b).

In order to obtain a frequency representation, Fast Fourier Transform (FFT) cannot be applied directly because it is an unevenly spaced signal, and it requires a prior transformation. Signal resampling by interpolation is the common strategy to address this problem. Once the spectrum is estimated, it is separated into frequency bands, which are related to the JMs and swallowing-regurgitation pauses during rumination.

\begin{figure}
       \centering
       \includegraphics[width=.8\textwidth]	{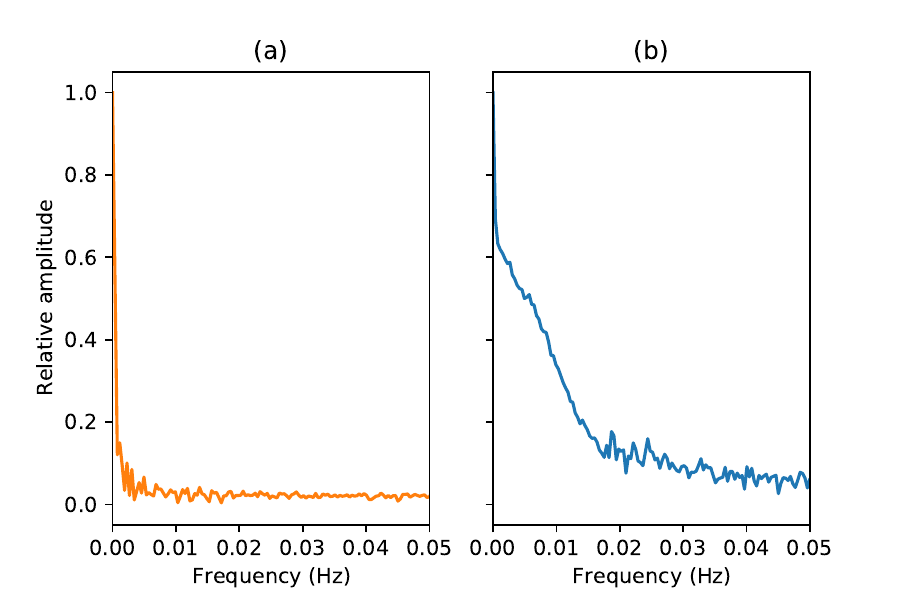}
       \caption{Spectrum of (a) grazing and (b) rumination tachograms.}
       \label{fig:spectTacog}
\end{figure}

Figure~\ref{fig:spectTacog} shows typical spectra of tachograms for grazing and rumination. The spectra show clear differences in the frequency band between 0~Hz and 0.02~Hz. These differences are caused by the swallowing-regurgitation pauses during rumination, whose periodicity is revealed by the low frequency components close to zero. Both spectra show high intensity in very low frequencies, which is related to the offset close to 1~s shown in both tachograms (Figure~\ref{fig:tachograms}). This contribution is not expected for activities other than grazing and rumination, where this kind of periodic events may not be present. For example, sip events are periodic but have a lower rate compared to chewing and biting JM \citep{galli2008Thesis}.

\subsubsection{Jitter variants}

\begin{figure}
       \centering
       \includegraphics[width=.9\textwidth]	{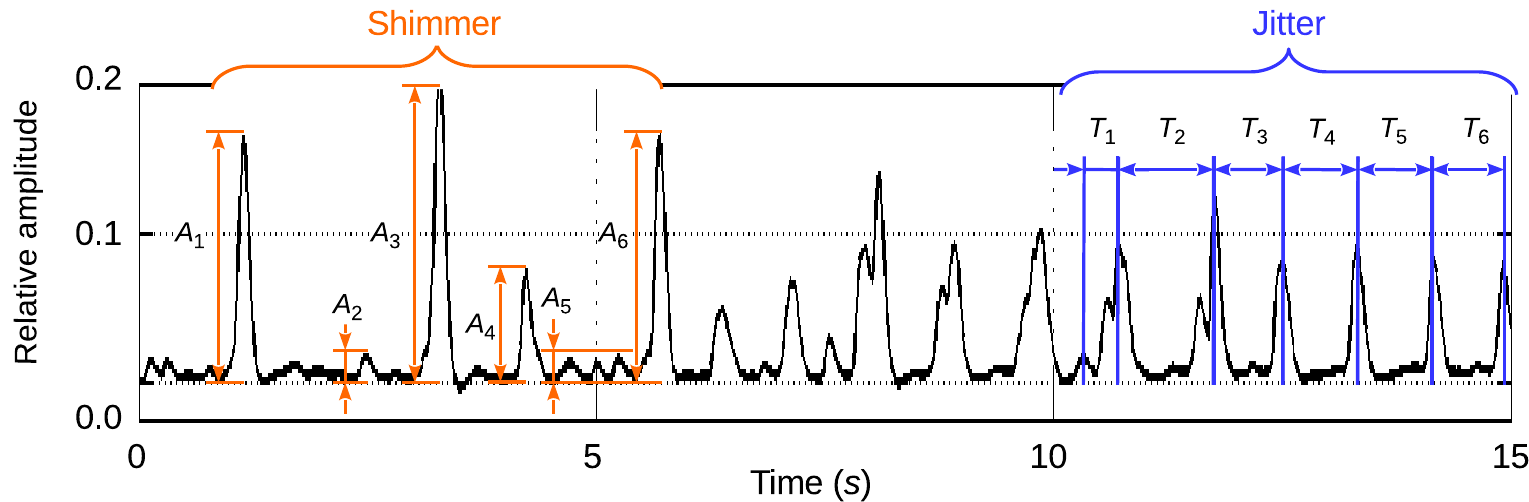}
       \caption{Representation of jitter and shimmer measures in audio envelope.}
       \label{fig:jitterAndShimmer}
\end{figure}

Jitter is the variation of time period or latency that occurs between consecutive JM~events (Figure~\ref{fig:jitterAndShimmer}). During grazing bites, chews and chew-bites occur with an inter-event duration period that varies around 1 s; greater inter-event time interval variation indicates greater jitter. During rumination, regular chews occur with an inter-event interval of lower variability compared to grazing, thereby with lower jitter. Other activities showing more irregular occurrence of detected events, will have greater jitter value compared to grazing and rumination.

There are alternative variants for the quantification of jitter:

\begin{itemize}
    \item {Absolute jitter ($jitter_{abs}$) - is the average absolute difference between consecutive inter-event periods within the given sound segment
    \begin{equation}
        jitter_{abs}=\frac{1}{N-1}\sum_{i=1}^{N-1}\left | T_i - T_{i+1} \right |
    \label{eq:1}
    \end{equation}
    where $T_i$ is the $i-th$ time period between two consecutive events and $N$ is the number of inter-event periods in the sound signal segment being analyzed.
    }
    
    \item {Relative jitter ($jitter_{rel}$) - is the average absolute difference between consecutive inter-event periods divided by the average inter-event period of the sound segment
    \begin{equation}
        jitter_{rel}=\frac{\frac{1}{N-1}\sum_{i=1}^{N-1}\left | T_i - T_{i+1} \right |}{\frac{1}{N}\sum_{i=1}^{N}T_i}
    \label{eq:2}
    \end{equation}
    }
    
    \item {PPQ5~jitter ($jitter_{ppq5}$) - is the five-point period perturbation quotient computed as the average absolute difference between an inter-event period and the average of that period and the four closest neighbor periods, divided by the average period of the sound segment
    
    \begin{equation}
    jitter_{ppq5}=\frac{\frac{1}{N-4}\sum_{i=3}^{N-2}\left | T_i - \frac{1}{5}(T_{i-2}+T_{i-1}+T_i+T_{i+1}+T_{i+2}) \right |}{\frac{1}{N}\sum_{i=1}^{N}T_i}
    \label{eq:3}
    \end{equation}
    }
\end{itemize}

The absolute jitter quantifies the variability of the inter-event periods in a sound segment without taking into account the average period. Thus, the absolute jitter loses sensitivity when the inter-event period changes. This problem is solved by the relative jitter, which scales the absolute jitter by the average period of the sound segment. It expresses the inter-event variability as a fraction of the average period, providing a good sensitivity to inter-event period changes. However, these alternatives (equations~\ref{eq:1} and~\ref{eq:2}) focus only on event variability, introducing a high variability in the jitter indexes. To incorporate information of event neighbors, the PPQ5~jitter is employed. This jitter variant considers the average of a period and the four nearest neighbor periods to compute the inter-event period information.

\subsubsection{Shimmer variants}

Shimmer is defined as the variation of peak to peak signal amplitude~(dB) between events (Figure~\ref{fig:jitterAndShimmer}). During grazing, bites, chews and chew-bites occur along a wide range of signal amplitudes; thus the shimmer between events has a high value. During rumination, rumination chews exhibit lower variation in signal amplitude, or have a lower shimmer value compared to grazing. Other activities have no regularity in their amplitude, showing a shimmer value higher than grazing and rumination (not shown).

There are alternative variants for the calculation of shimmer:

\begin{itemize}
    \item {Absolute shimmer ($shimmer_{abs}$) - is the variability of the peak-to-peak amplitude in decibels, which is computed as the average of absolute base-10 logarithm of the ratio between the amplitudes of consecutive events, multiplied by 20,
    
    \begin{equation}
        shimmer_{abs}=\frac{1}{N-1}\sum_{i=1}^{N-1}\left | 20log (A_{i+1}/A_i) \right |,
    \label{eq:4}
    \end{equation}
    
    where $A_i$ is the $i-th$ extracted peak-to-peak amplitude and $N$ is the number of detected events in the sound segment.
    }
    
    \item {Relative shimmer ($shimmer_{rel}$) - is the average absolute difference between amplitudes of consecutive events divided by the average amplitude of events in the sound segment,
    
    \begin{equation}
        shimmer_{rel}=\frac{\frac{1}{N-1}\sum_{i=1}^{N-1}\left | 20log (A_{i+1}/A_i) \right |}{\frac{1}{N}\sum_{i=1}^{N}A_i}
    \label{eq:5}
    \end{equation}
    }
    
    \item {APQ5~shimmer ($shimmer_{apq5}$) - is the five-point amplitude perturbation quotient, or the average absolute difference between the amplitude of an event and the average of the amplitudes of that event and the four closest neighbors events, divided by the average amplitude of events in the sound segment. 
    
    \begin{equation}
        shimmer_{apq5}=\frac{\frac{1}{N-4}\sum_{i=3}^{N-2}\left | A_i - \frac{1}{5}(A_{i-2}+A_{i-1}+A_i+A_{i+1}+A_{i+2}) \right |}{\frac{1}{N}\sum_{i=1}^{N}A_i}
    \label{eq:6}
    \end{equation}
    }
\end{itemize}

Like jitter, the absolute shimmer quantifies the amplitude variability between consecutive events in a sound segment without taking into account the average amplitude, thus it loses sensitivity when the average amplitude changes. This problem is solved using the relative shimmer, which scales the absolute shimmer by the average amplitude of the sound segment. The relative variant expresses the inter-event amplitude variability as a fraction of the average amplitude, thereby providing a good sensitivity to inter-event amplitude changes. However, these shimmer variants (equations~\ref{eq:4} and~\ref{eq:5}) focus only on event variability, introducing a high variability in the shimmer indexes. To incorporate information of event neighbors the APQ5~shimmer is employed. This shimmer variant considers the average of amplitude of a given event and the four closest neighbors to compute this inter-event amplitude information.

\section{Experimental setup}

\subsection{Variants proposed}

Different combinations of the proposed algorithm were analyzed. Table~\ref{tab:features} shows the features used by the former reference method (\mbox{BUFAR}), the proposed method called JM segment-based Foraging Activity Recognizer (\mbox{JMFAR}), and the variants analyzed. Two aspects were taken into account: (i)~the recognition performance and (ii)~the computational cost. The variants analyzed were:

\begin{itemize}
    \item {JMFAR: This variant is the proposed method considering all new features.}
    \item {\mbox{JMFAR-sel}: This variant considers the resulting features after applying a feature selection method on the set of {JM-segment features}.}
    \item {\mbox{JMFAR-ns}: This variant does not consider the spectral information (f13, f23 and f24) in order to reduce the computational cost and memory requirements of the algorithm 
    (25\% of reduction). This is desired to reduce the power consumption and increase the battery-life in wearable devices.
    }
\end{itemize}

For all the proposed methods features related to JM~classification (f1, f2 and f3) were excluded.

\begin{table}
\centering
\caption{Features used by \mbox{BUFAR}, the proposed method \mbox{JMFAR} and its variants.}
\label{tab:features}
\resizebox{\textwidth}{!}{\begin{tabular}{lcccccccccccccccccccccccc}

Method      & \multicolumn{24}{c}{Features}                                                                                                        \\ \hline
            & f1 & f2 & f3 & f4 & f5 & f6 & f7 & f8 & f9 & f10 & f11 & f12 & f13 & f14 & f15 & f16 & f17 & f18 & f19 & f20 & f21 & f22 & f23 & f24 \\ \hline
\mbox{BUFAR}       &  $\bullet$  &  $\bullet$  &  $\bullet$  &  $\bullet$  &    &    &    &    &    &     &     &     &     &     &     &     &     &     &     &     &     &     &     &     \\ \hline
{\mbox{JMFAR}}     &    &    &    &  $\bullet$  &  $\bullet$  &  $\bullet$  &  $\bullet$  &  $\bullet$  &  $\bullet$  &   $\bullet$  &  $\bullet$   &   $\bullet$  &  $\bullet$   &  $\bullet$   &  $\bullet$   &   $\bullet$  &  $\bullet$   &   $\bullet$  &  $\bullet$   &  $\bullet$   &   $\bullet$  &   $\bullet$  &  $\bullet$   &   $\bullet$  \\ \hline
{\mbox{JMFAR-sel}} &    &    &    &    &    &  $\bullet$  &  $\bullet$  &  $\bullet$  &    &     &  $\bullet$   &   $\bullet$  &   $\bullet$  &     &  $\bullet$   &     &   $\bullet$  &   $\bullet$  &  $\bullet$   &  $\bullet$   &     &   $\bullet$  &     &     \\ \hline
{\mbox{JMFAR-ns}}  &    &    &    &  $\bullet$  &  $\bullet$  &  $\bullet$  &  $\bullet$  &   $\bullet$ &  $\bullet$  &   $\bullet$  &   $\bullet$  &   $\bullet$  &     &  $\bullet$   &   $\bullet$  &  $\bullet$   &  $\bullet$   &  $\bullet$   &   $\bullet$  &  $\bullet$   &  $\bullet$   &  $\bullet$   &     &     \\ \hline
\end{tabular}}
\end{table}

The feature selection was performed with a wrapper method using a genetic algorithm as a search method. 
{As the internal classifier of the wrapper, a multilayer perceptron similar to the external classifier was used. The parameters used were a mutation probability of 0.2, a crossover probability of 0.5, a population size of 50 individuals and a maximum of 30 generations \citep{kohavi1997wrappers}}. For this process, a 5-fold scheme on the training set was used. The accuracy was used as the scoring measure in the selection process. The features that were selected at least in two folds were included.

For this study, a set of 24~h of recordings belonging to DbM (32.0\% grazing, 50.3\% rumination and 17.6\% of other activities)
were used to train and optimize parameters and hyper-parameters of the activity classifier and they were never used again. Classifiers were trained\footnote{This stage was implemented in python using the scikit-learn package.} following a 5-fold cross validation on the training set. Finally, the test results were obtained independently from a separate test set of 137~h of recordings belonging to DbM (48.4\% grazing, 38.6\% rumination and 13.0\% of other activities) and another test set of 30~h of recordings belonging to DbZ (38.3\% grazing, 23.7\% rumination and 38.0\% of other activities). All recordings were selected taking care that they correspond to a free ranging environment (i.e. outside of the feeding barn).
Evaluation on DbM indicated the performance of the algorithm variants in similar train/test conditions (animals, diet and field conditions, among others), whilst test on DbZ was used to evaluate the generalization ability of the JMFAR on signals recorded under different experimental conditions (animals, diet and field conditions, among others).

\subsection{Performance assessment}

The weighted average F1-score \citep{blair1979information} was computed as a weighted metric for the precision and recall of methods. This metric summarizes the overall performance of methods and for their specificity for classification of grazing and rumination. In a previous work the authors presented results at the frame and at the block-level using spider-plots \citep{chelotti2020online}. However, in order to obtain each individual measure per signal, the proportion and duration of each activity were assumed to be equal. Although the information in the spider-plots is useful, in this work the actual length of time is taken into account and presented in a summarized and more accurate form. The F1-scores were computed using sequences of 1~s frames for the analysis. For each audio signal three F1-scores were computed, one for rumination, one for grazing, and one average score weighed by the number of frames for each activity. F1-scores range between 0.0 and 1.0 and values close to 1.0 indicate better activity classification performance.

The analysis of each audio signal provided a list of blocks where the start, end, and type of activity are specified. These blocks were partitioned to 1~s non-overlapping frames sequences. Due to the fact that activity bouts last several minutes or even hours, using 1~s frames provides a very detailed resolution of results. Further, any modification of the limits of an activity block below 1~s can be discarded from the point of view of the recognition problem. This transformation was applied both to the algorithm output and ground truth of each signal. After this transformation there are corresponding pairs of sequences of labeled frames for the ground truth and each analyzed algorithm. Thus, conventional metrics for classification problems can be used.

\section{Results}

\subsection{Comparative performance of \mbox{BUFAR} and {\mbox{JMFAR}} methods}

\begin{figure}
       \centering
       \includegraphics[width=.75\textwidth]	{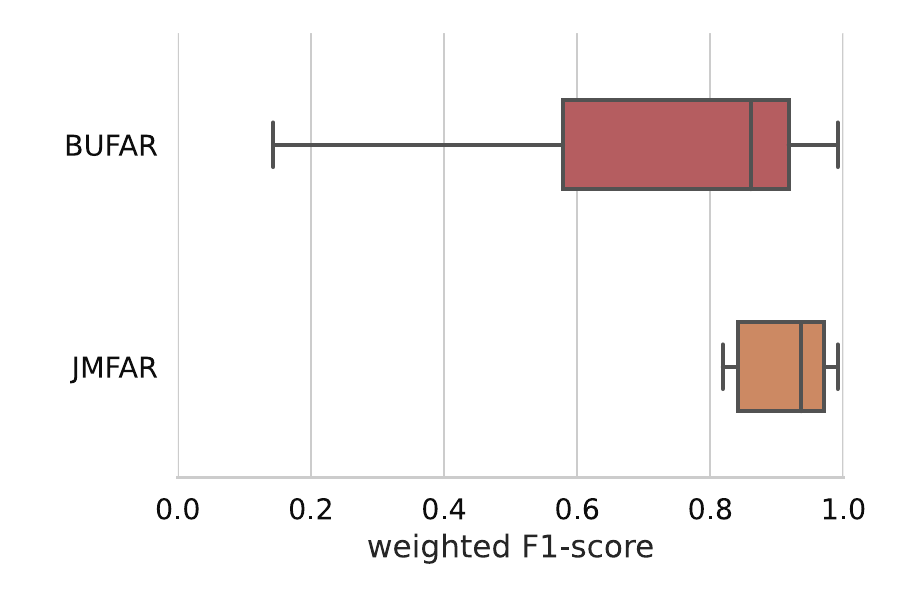}
       \caption{Weighted average of frame F1-score computed in the DbM database.}
       \label{fig:BoxPlotWhisker}
\end{figure}

The boxplots of Figure~\ref{fig:BoxPlotWhisker} show the weighted F1-score for \mbox{BUFAR} and {\mbox{JMFAR}} algorithms. Each data point considered for the graphic corresponds to 
an analyzed audio signal belonging to the DbM database. The interquartile range for F1-scores of \mbox{BUFAR} spans from 0.55 to 0.91 and has a median of 0.86. By contrast, the {\mbox{JMFAR}} algorithm shows a narrower range and distribution of F1-scores. The interquartile range of {\mbox{JMFAR}} spans from 0.83 to 0.97 and has a median of 0.93. Score differences between \mbox{BUFAR} and {\mbox{JMFAR}} have shown to be significant (p=0.012) according to a Wilcoxon signed-rank test comparison \citep{wilcoxon1947probability}.

\begin{figure}
\centering
\begin{subfigure}{.5\textwidth}
  \centering
  \includegraphics[width=0.9\linewidth]{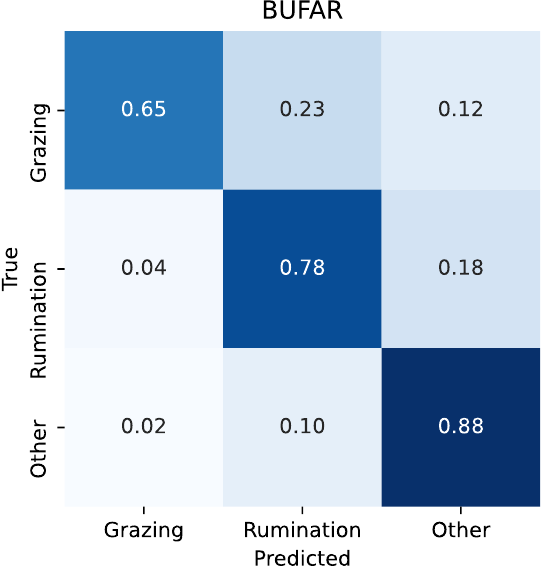}
\end{subfigure}%
\begin{subfigure}{.5\textwidth}
  \centering
  \includegraphics[width=0.9\linewidth]{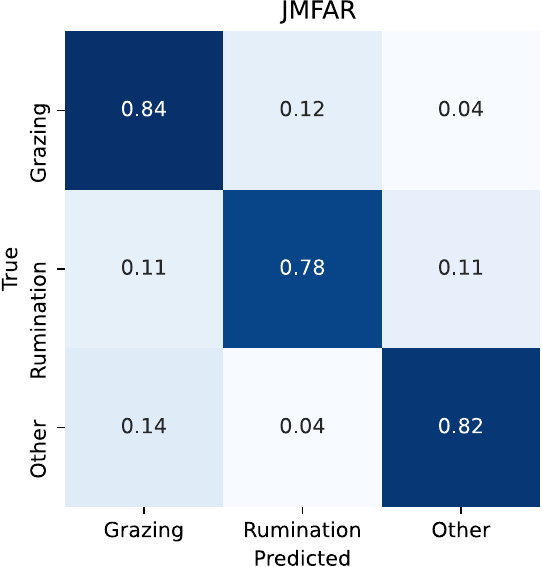}
\end{subfigure}
\caption{Confusion matrices for the \mbox{BUFAR} and \mbox{JMFAR} algorithms. The improvement in performance for grazing recognition is clear.}
\label{fig:BUFARVsJMFAR}
\end{figure}

Figure~\ref{fig:BUFARVsJMFAR} shows the confusion matrices for the \mbox{BUFAR} and {\mbox{JMFAR}}. Rows represent frames and their actual activity label, whereas columns represent the frames and activity label as they were assigned by the algorithm. Each cell [i,~j] provides the proportions of frames from the activity class~i that were categorized as activity~j by the algorithm. The cells along the diagonal indicate the correct classifications. A classifier that provides correct labels for each frame would generate a unit matrix. Off diagonal cells represent different kinds of errors for incorrectly classified frames. Matrix values range from 0.0 (white) to 1.0 (dark blue). The matrix for \mbox{BUFAR} shows that 65\% of grazing and 78\% of rumination frames were correctly classified. Further, 23\% of grazing frames were incorrectly labeled as rumination. 
On undetected frames,
the underestimation of activities was 12\% for grazing and 18\% for rumination.
{\mbox{JMFAR}} confusion matrix shows better results for the activities of interest. Recognition was correct for 84\% of grazing and 78\% for rumination. Confusions among rumination and grazing were similar (11 and 12\%) and undetected frames were much lower, 4\% for grazing, and 11\% for rumination.

Previous results can be interpreted temporally as a function of sound signal length. For example, if a 180~min audio signal corresponds to 60~min of grazing, 60~min of rumination and 60~min to other activities, classification results would have been as follows. The \mbox{BUFAR} algorithm would have predicted 42.6~min of grazing and 66.6~min of rumination, with 13.8~min of true grazing activity mislabeled as rumination and 2.4~min of true rumination mislabeled as grazing. Similarly, classifications by {\mbox{JMFAR}} would have improved to 65.4~min of grazing and 56.4~min of rumination, with 7.2~min of true grazing mislabeled as rumination and 6.6~min of true rumination mislabeled as grazing. Additional spider plots for classifications of grazing and rumination by \mbox{BUFAR} and \mbox{JMFAR} algorithms are available in the supplementary material. Plots use the same metrics as in \citet{chelotti2020online}, allowing for direct study comparisons of methods.

\begin{figure}
       \centering
       \includegraphics[width=.75\textwidth]	{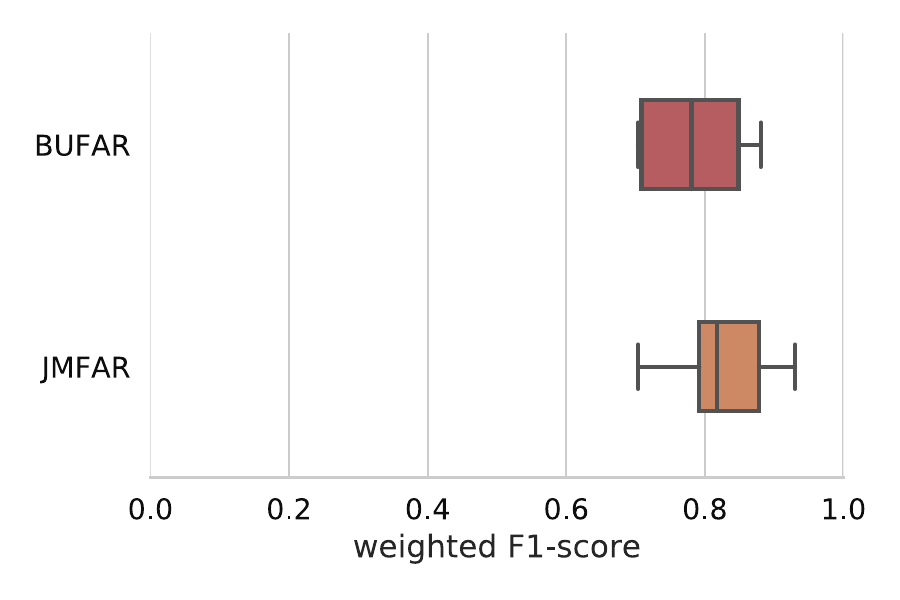}
       \caption{Weighted average of frame F1-score computed in the DbZ database.}
       \label{fig:boxplots_db2}
\end{figure}

Finally, in order to evaluate the robustness of the algorithm, the model originally trained on signals belonging to the DbM was validated on signals belonging to the DbZ. The boxplots in Figure~\ref{fig:boxplots_db2} show the weighted F1-score for BUFAR and JMFAR algorithms evaluated on DbZ. Each data point considered for the graphic corresponds to an analyzed audio signal belonging to the DbZ database. The interquartile range for F1-scores of BUFAR spans from 0.71 to 0.85 and has a median of 0.78. By contrast, the JMFAR algorithm shows a narrower range and distribution of F1-scores. The interquartile range of JMFAR spans from 0.79 to 0.88 and has a median of 0.82. Score differences between BUFAR and JMFAR have shown to be significant (p=0.031) according to a Wilcoxon signed-rank test comparison \citep{wilcoxon1947probability}.

\subsection{Variants of the proposed algorithm}

\begin{figure}
\centering
\begin{subfigure}{.5\textwidth}
  \centering
  \includegraphics[width=0.9\linewidth]{Figures/JMFAR_matrix.pdf}
\end{subfigure}%
\begin{subfigure}{.5\textwidth}
  \centering
  \includegraphics[width=0.9\linewidth]{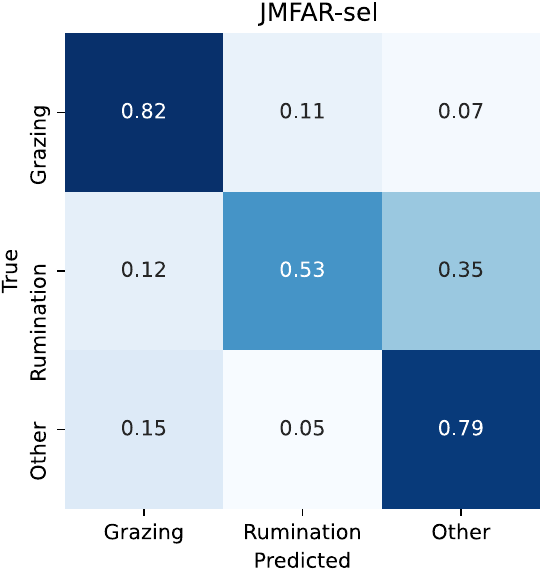}
\end{subfigure}
\begin{subfigure}{.5\textwidth}
  \centering
  \includegraphics[width=0.9\linewidth]{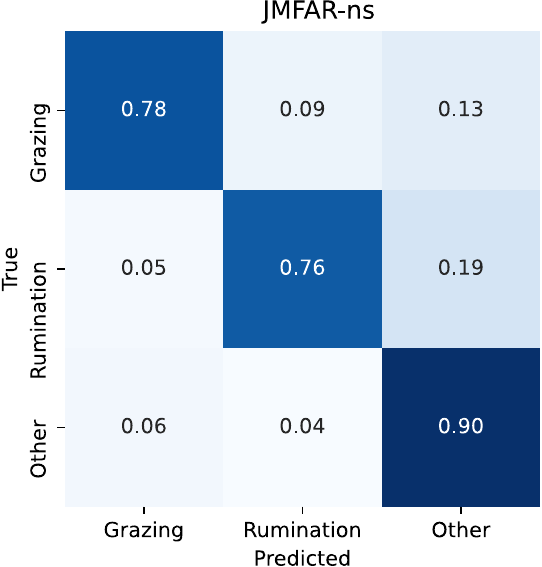}
\end{subfigure}
\caption{Confusion matrices of the algorithm variants analyzed.}
\label{fig:variant_JMFAR}
\end{figure}

Figure~\ref{fig:variant_JMFAR} shows the confusion matrices for {\mbox{JMFAR}} and variants {\mbox{JMFAR-sel}} and \mbox{JMFAR-ns}. The \mbox{JMFAR-sel} variant only uses 12 of the proposed JM-segment features, allowing savings on computational costs. However, benefits are quickly offset by weaker classification results. Correct classifications were 82\% for grazing and only 53\% for rumination, with a 47\% of undetected rumination frames. The \mbox{JMFAR-ns} variant achieves satisfactory results for grazing and rumination (76\% and 78\%, respectively), with a lower misclassification between these activities (9\% and 5\% respectively). This variant has lower computational cost and buffer memory size, since spectral features were excluded from classification analyses. 

A summary of the average performance and computational costs for \mbox{JMFAR} and variants are presented in Table~\ref{tab:performance}. 
The \mbox{JMFAR} improves the previous recognition accuracy of \mbox{BUFAR} by 9.5\%, with a computational cost in the same order of magnitude.
In general the use of fewer classification features in \mbox{JMFAR-sel} and \mbox{JMFAR-ns} leads to a deterioration of the recognition accuracy compared to \mbox{JMFAR}. Remarkable is the deterioration of the recognition accuracy of rumination for \mbox{JMFAR-sel} (25\%), despite the improvement in computational cost. The average decline of recognition accuracy for \mbox{JMFAR-sel} was 13.5\% compared with \mbox{JMFAR} and 4\% compared to \mbox{BUFAR}, which occurred without incurring major alteration of computational costs. The \mbox{JMFAR-ns} also shows a decline in recognition accuracy compared to \mbox{JMFAR} (this decline was 6\% for grazing, 2\% for rumination), and an improvement of recognition accuracy compared to \mbox{BUFAR} (the improvement was 13\% for grazing, and 2\% for rumination). The computational cost was numerically lower for \mbox{JMFAR-ns} than \mbox{BUFAR}.

\begin{table}
\caption{Overall performance and computational cost of \mbox{BUFAR} and \mbox{JMFAR} variants.}
\label{tab:performance}
\centering
\footnotesize
\begin{tabular}{lcc}
\hline
\textbf{Method} & \begin{tabular}[c]{@{}c@{}}\textbf{Average accuracy}\\ (\%)\end{tabular} & \begin{tabular}[c]{@{}c@{}}\textbf{Computational cost}\\ (operations/s)\end{tabular} \\ \hline
\mbox{BUFAR}           & 71.5                                                            & 37,966                                                                      \\ \hline
{\mbox{JMFAR}}         & 81.0                                                            & 50,445                                                                      \\ \hline
{\mbox{JMFAR-sel}}     & 67.5                                                            & 43,736                                                                      \\ \hline
{\mbox{JMFAR-ns}}      & 77.0                                                            & 37,645                                                                      \\ \hline
\end{tabular}
\end{table}

\section{Discussion}

The individualized monitoring of grazing and rumination patterns could offer insights on sward–animal interactions affecting daily herbage intake, thereby providing unprecedented opportunities to optimize grazing management, enhance animal production or meet individualized requirements for animal nutrition, health, well-being and welfare \citep{beauchemin2018invited}. This study is limited to free grazing environment. Typical indoor scenarios, as a barn, could present additional challenges to the proposed method due to environmental noises. In addition, practical factors such as device placement, bumps and splashes could affect the performance of the system. These specific situations should be addressed in future studies.

The acoustic method has some advantages over pressure sensors and accelerometers. The sound
quality obtained by placing a microphone on the
forehead of the animal enabled algorithm to very
precisely discriminate the sounds associated with
diverse jaw movements during grazing and rumination.
Moreover, this precise detection capability enables to further expand the method to estimate the dry matter intake, to recognize the ingested forage species and also to obtain information about the surrounding environment \citep{galli2020discriminative, li2021}.

In this study the reliability and accuracy of applying on-line acoustic methods to monitor grazing and rumination patterns were assessed and compared. The proposed system was built using signal processing and machine learning concepts. Repositories of acoustic data collected with lactating dairy cows on a robotic milking facility were used for the process. Overall, increasing the number of temporal, statistical and spectral features in the classifier improved the accuracy of algorithms, whereas the inclusion of greater number of features and classification steps slightly increased computational costs compared to previous algorithms for operation on-line (i.e. in portable embedded systems). 

In this work, different sets of JM-segment features were evaluated for incorporation in a foraging activity recognizer called \mbox{JMFAR}. The proposed \mbox{JMFAR} algorithm showed better overall performance compared to \mbox{BUFAR} \citep{chelotti2020online}, a former algorithm developed on a bottom-up approach for recognition of foraging activities in cattle (Figures~\ref{fig:BoxPlotWhisker} and~\ref{fig:BUFARVsJMFAR}). \mbox{JMFAR} achieved 84\% and 78\% for grazing and rumination recognition, respectively and a misclassification between grazing and rumination frames of 12\% and 11\%. Furthermore, the recognition of grazing was improved from 65\% to 84\% compared to the former algorithm \mbox{BUFAR}. This is an important achievement because \mbox{BUFAR} can misclassify about 35\% of true grazing frames either as rumination (23\%) or other non-grazing activity frames (12\%). A less significant misclassification problem of \mbox{BUFAR} could be seen for rumination, where 18\% of true rumination frames could be incorrectly assigned to other non-grazing activities. Hypothetically, the issue of misclassifications by \mbox{BUFAR} could be associated with a limited availability of statistical information on JM in the classifier. In \mbox{BUFAR} most of the activity classification lies on the accuracy of JM classification. In this sense, the new algorithm \mbox{JMFAR} overcomes these limitations by using new {JM-segment features} of acoustic data, which makes the new {activity} classifier fully independent of the {JM-classification}. Indeed, \mbox{JMFAR} bridges completely the step for JM~classification previously addressed by \mbox{BUFAR}. In \mbox{JMFAR} and the variants tested, new arrays and variants of JM-segment features of sound data such as the jitter, the shimmer and the sound spectra, are increasingly relevant. This sequence view rather than a simple group view of detected JM, clearly makes the difference in the quality of activity classifications. The sequence patterns of acoustic features and their differences between grazing and rumination are the key distinctive elements between these two activities and for other non-grazing and non-rumination activities.

The robustness of the algorithm was also evaluated (Figure~\ref{fig:boxplots_db2}). The new algorithm, originally developed using signals belonging to the DbM, was tested on signals belonging to the DbZ. Performances of BUFAR and JMFAR algorithms under this schema were slightly lower than the obtained from testing on signals belonging to the DbM. Again, JMFAR improved the results for BUFAR in terms of F1 score (medians of 0.82 and 0.78 respectively). This external validation approach represents a challenging scenario due to databases are generally obtained under different experimental conditions.

Among the variants tested, the \mbox{JMFAR-sel} which relied on a supervised selection for JM-segment features, achieved the most modest results (Figure~\ref{fig:variant_JMFAR}). This shows that in order to reduce the computational cost, retaining only the most discriminative features, relevant information is lost. The overall performance of \mbox{JMFAR-ns}, a variant that excluded spectral features in the classifier, showed a slightly lower accuracy compared to \mbox{JMFAR} (77\% vs 81\% from Table~\ref{tab:performance}), although computational costs were reduced in \mbox{JMFAR-ns}.
This variant achieved the lowest confusion of classification between grazing and rumination frames (which were 9\% and 5\%) yet it had a tendency to underestimate grazing and rumination by 13\% and 19\%, respectively. Hypothetically, this underestimation could be due to an exclusion of the 1.0-1.5~Hz energy band from the analysis, which is associated with regularly performing JM, increasingly significant in grazing and rumination chewing. Interesting results were observed when both JM-classification and JM-segment features were used together. In this case, the overall performance was 84\% and 81\% for grazing and rumination, respectively. This implies an overall improvement of 3\% for rumination in comparison with \mbox{BUFAR}, which could be due to a lower confusion of classification between grazing and rumination. The addition of JM~proportional information, in particular the addition of proportions of chews during rumination, improved the recognition performance of rumination and overall. This findings support the hypothesis that greater classification accuracy is the result of distinct contributions aggregating across sets of discrimination features (i.e. statistical, temporal or spectral), thereby improving the overall activity recognition process.

In practice, for on-line analyses tallying a daily monitoring with 8.00~h of true grazing, \mbox{BUFAR} would have indicated the equivalent median of 9~h~28~min (interquartile range of 8~h~13~min - 12~h~14~min), while \mbox{JMFAR} would indicate the equivalent median of 7~h~59~min (interquartile range of 7~h~19~min - 8~h~35~min) of classified grazing. Assuming literature values of 1.5~kg/h of dry matter consumption \citep{Galli2018-zq}, previous values correspond to an approximate consumption with a median of 14.2~kg (interquartile range 12.3~-~18.3~kg) and 12.0~kg (interquartile range 11.0~-~12.8~kg) for \mbox{BUFAR} and \mbox{JMFAR} respectively. Certainly, additional research would be necessary to evaluate the application of acoustic methods for estimation of herbage dry matter intake. As the pioneer work by \citet{laca2000acoustic} has provided the proof of concept for the use of ingestive sounds for prediction of herbage dry matter intake of sheep \citep{galli2011acoustic}, cattle \citep{Galli2018-zq}, and potentially other herbivore species. For example chewing and biting sounds apparently enclose statistical, energy and possibly spectral information that could be used as metrics for accurate estimations of dry matter intake (CV~17\%) of cows and even when those cows are confronted with forages that have varying physical and chemical composition yet comparable sound signal properties and metrics. Despite progresses, further work is still necessary to apply extractions of acoustic and information from \mbox{JMFAR} and variants for scalable predictions of herbage dry matter intake across animals.

In this study, \mbox{JMFAR} and variants explored likely trade-offs between classification performance and the associated computational costs (Table~\ref{tab:performance}). The former JMFAR, considering all new JM-segment features, requires 50,445 operations per second (ops/s). There are two variants that have attempted to minimize the computational cost. The \mbox{JMFAR-sel}, reduced the reference computational cost by 13.3\% at expense of reducing the overall performance from 81.0\% to 67.5\%. The \mbox{JMFAR-ns} variant, required 37,645~ops/s, and yet it has reduced costs by 25.4\% the performance was reduced from 81.0\% to 77.0\%, despite the fact that this variant had better performance and lower computational cost compared to the preceding algorithm \mbox{BUFAR}. In the \mbox{JMFAR-ns} variant the classifier input set was reduced to 17  features in the time domain, avoiding performing fast Fourier transformations, which are not always feasible to realize in low power microcontrollers. 
The exclusion of spectral features reduced 99.9\% the computational cost associated to conform the set of activity features.
Thus, fewer features included in this variant and avoidance for prior classification of JM would have been responsible for all computational cost improvements.
A detailed analysis of the computational cost is included in the~\ref{appendix_A}.
The requirement for random access memory (RAM) is another key feature for the on-line implementation of algorithms in portable embedded systems. As a reference, a simple 8-bit data representation with storage rate of 100~Hz, would have required 30~kB of RAM to store a segment of a given sound envelope. Furthermore, the implementation of \mbox{JMFAR} and \mbox{JMFAR-sel} variants would have required an addition of 30~kB of RAM to store the spectral representation of the sound envelope. Thus, after consideration of a proportional consumption of power, the \mbox{JMFAR-ns} variant is suggested as the most appropriate method for implementation in a low power microcontroller-based system. 

This work was performed with the same number of animals as in previous
studies \citep{Vanrell2018-zj, chelotti2020online}, with a total of 250 hours analyzed. This is a first step, in
which we test the performance of the system with a homogeneous group of
animals. Future work will contemplate the expansion of these
results to a much wider range, including a larger number of animals and
incorporating different frames, sexes, breeds and age.
The level of variation between cows in this experiment should not represent a
problem for the application of the method, in fact but, in principle it could be
extrapolated to groups of cows homogeneous in breed, size and age. The
degree of variation between individuals is important because it will determine
the generality of predictions based on chewing sounds. Chewing activity could
present individual differences due to the influence of sex, body
weight, physiological state and age, but this should not necessarily cause
differences in ingestive or ruminant chewing.
The cows used in this trial were similar in live weight, age and size, but could
differ in the size and structure of their teeth, as well as the size and shape of
their heads that produced different chewing sounds. In the worst case scenario, these uncontrolled
sources of variation could suggest that a calibration of the recording equipment for each individual might be required when the animals are not uniform.

\section{Conclusions}

A novel acoustic method for the recognition and estimation of grazing and rumination bouts has been proposed. The proposed \mbox{JMFAR} is based on JM-segment features, i.e. that the computation of each feature is independent of the JM~identification. The proposed features were: statistics of JM~amplitudes and JM~durations, spectral features of the sound envelope, JM~jitter and JM~shimmer variants, temporal and spectral features of the JM~tachogram. The algorithm has shown an improved performance in comparison with the \mbox{BUFAR method based on JM-classification features}. Variants relying on maximum numbers for JM-segment features achieved the best overall classification results with slight trade-offs on computational costs, and memory requirements for use in portable embedded systems. Among the variants analyzed, the one that ignores spectral information showed the best trade-off between recognition performance and computational cost.

\section*{Acknowledgments}

This study has been funded by Universidad Nacional del Litoral, PACT CAID 2011–525, Universidad Nacional de Rosario, projects 2013-AGR216, 2016-AGR266 and 80020180300053UR, Agencia Santafesina de Ciencia, Tecnología e Innovación (ASACTEI), project IO-2018-00082, Consejo Nacional de Investigaciones Científicas y Técnicas (CONICET), project 2017-PUE-sinc(i). Also, this study was funded under the USDA-NIFA MICL0222 and MICL0406 projects, and direct support from AgBioResearch-MSU. Authors wish to thank the insightful help and dedication by the KBS Robotic Dairy Farm staff for their assistance and support during the completion of this study.

\section*{AUTHORSHIP STATEMENT}

JOC, SRV, JRG, SAU, DHM, LLG and HLR participated in conceptualization; JOC, SRV, JRG, AMP and SA participated in the data curation; JOC, SRV and LSMR participated in the formal analysis; JOC, SRV, JRG, AMP, SAU, DHM, LLG and HLR participated in investigation and methodology stages; JOC and SRV participated in validation and visualization stages; SRV, JRG, SAU, DHM, LLG, HLR participated in the funding acquisition; JOC, JRG, DHM, LLG and HLR participated in project administration; All the authors contributed to the writing and reviewing of the original draft; All the authors reviewed and approved the manuscript.

\appendix
\section{Computational cost}
\label{appendix_A}
The computation cost of the \mbox{JMFAR} variants depends on the sampling frequency, the considered segment length and the number of features used by the activity classifier. In this analysis a sampling frequency of 2~kHz, a segment length of 5~min and two JM per second were considered in order to get a simple comparison with other methods. Worst-case scenarios were considered for each stage in order to get a theoretical upper bound. The required number of operations per stage of computation for \mbox{JMFAR} was:

\begin{enumerate}
    \item {Signal pre-processing: A detrending filter requires 5~operations per signal sample. Then, 10,000~ops/s are required.}
    \item {JM~detection: 27,200~ops/s are required to compute the subsampled envelope, generate a time-varying threshold and detect the jaw movement.}
    \item {JM~feature extraction: This task involves 300~ops/s.}
    \item {Buffering: 6~ops/s are required to store the timestamp, and the amplitude and duration features.}
    \item {Activity feature extraction: the worst-case scenario for activity feature extraction considers 2 event/s, which is equivalent to 600 events per segment.
    \begin{enumerate}
        \item {JM~rate (f4): this feature requires only 1 op/segment.}
        \item {JM~statistics:
        \begin{enumerate}
            \item {mean amplitude (f5) and mean duration (f9): each feature requires 600~ops/segment.}
            \item {standard deviation amplitude (f6) and standard deviation duration (f10): 1801~ops/segment per feature are required.}
            \item {skewness amplitude (f7), skewness duration (f11): compute each feature requires 2,401~ops/segment.}
            \item {kurtosis amplitude (f8) and kurtosis duration (f12): compute each feature requires 3,001~ops/segment.}
        \end{enumerate}
        \item {Spectral feature of the envelope signal (f13): compute the energy in the 1.0-1.5~Hz band requires 1,830,252~ops/segment.}
        \item {JM~jitter and shimmer variants: 
        \begin{enumerate}
            \item {absolute variants: absolute jitter (f14) requires 2396~ops/segment and compute the absolute shimmer (f18) requires 2398~ops/segment.}
            \item {relative variants: absolute jitter (f14) and absolute shimmer (f18) are used to compute relative jitter (f15) and relative shimmer (f19), respectively. Each feature requires 601 extra~ops/segment.}
            \item {perturbation quotient variants: partial calculations of feature (f15) and partial calculations of feature (f18) are used to calculate the PPQ5~jitter (f16) and the APQ5~shimmer (f20), respectively. Additional 4770~ops/segment are required per feature computed.}
            \item {standard deviation variants: absolute jitter (f14) is used to calculate the standard deviation of jitter (f17). Additional 2,997~ops/segment are required. Similarly, the feature (f15) is used to compute the standard deviation of shimmer (f21) involving 4,195~ops/segment.}
        \end{enumerate}
        \item {Temporal feature of the JM~tachogram (f22): worst-case scenario requires 1,201~ops/segment.}
        \item {Spectral features extraction of the JM~tachogram requires both an interpolation, a frequency domain transform, and an energy computation involving 2,009,479~ops/segment. Additionally, the computation of energy in the 0.017-0.020 Hz (f23) and 0.00-0.02 Hz (f24) bands requires 5 and 53~ops/segment, respectively.}}
        }
    \end{enumerate}
    \item {Activity classifier (MLP): considering all extracted features (21) and the largest number of neurons in the hidden layer (20), the activity classification task requires 2,279~ops/segment.}}
\end{enumerate}

The overall computation cost of \mbox{JMFAR} is: 37,506~ops/s + 3,881,604~ops/segment. To compare the algorithm with an on-line method with the same block length, the computational cost can be estimated at 50,445~ops/s.


\newpage
\bibliographystyle{elsarticle-harv}
\bibliography{references.bib}

\begin{thebibliography}{57}
\expandafter\ifx\csname natexlab\endcsname\relax\def\natexlab#1{#1}\fi
\providecommand{\url}[1]{\texttt{#1}}
\providecommand{\href}[2]{#2}
\providecommand{\path}[1]{#1}
\providecommand{\DOIprefix}{doi:}
\providecommand{\ArXivprefix}{arXiv:}
\providecommand{\URLprefix}{URL: }
\providecommand{\Pubmedprefix}{pmid:}
\providecommand{\doi}[1]{\href{http://dx.doi.org/#1}{\path{#1}}}
\providecommand{\Pubmed}[1]{\href{pmid:#1}{\path{#1}}}
\providecommand{\bibinfo}[2]{#2}
\ifx\xfnm\relax \def\xfnm[#1]{\unskip,\space#1}\fi
\bibitem[{Andriamandroso et~al.(2016)Andriamandroso, Bindelle, Mercatoris and
  Lebeau}]{Andriamandroso2016-lr}
\bibinfo{author}{Andriamandroso, A.}, \bibinfo{author}{Bindelle, J.},
  \bibinfo{author}{Mercatoris, B.}, \bibinfo{author}{Lebeau, F.},
  \bibinfo{year}{2016}.
\newblock \bibinfo{title}{A review on the use of sensors to monitor cattle jaw
  movements and behavior when grazing}.
\newblock \bibinfo{journal}{Biotechnologie, Agronomie, Soci{\'e}t{\'e} et
  Environnement} \bibinfo{volume}{20}.
\bibitem[{Arablouei et~al.(2021)Arablouei, Currie, Kusy, Ingham, Greenwood and
  Bishop-Hurley}]{arablouei2021situ}
\bibinfo{author}{Arablouei, R.}, \bibinfo{author}{Currie, L.},
  \bibinfo{author}{Kusy, B.}, \bibinfo{author}{Ingham, A.},
  \bibinfo{author}{Greenwood, P.L.}, \bibinfo{author}{Bishop-Hurley, G.},
  \bibinfo{year}{2021}.
\newblock \bibinfo{title}{In-situ classification of cattle behavior using
  accelerometry data}.
\newblock \bibinfo{journal}{Computers and Electronics in Agriculture}
  \bibinfo{volume}{183}, \bibinfo{pages}{106045}.
\bibitem[{Bailey et~al.(1996)Bailey, Gross, Laca, Rittenhouse, Coughenour,
  Swift and Sims}]{bailey1996mechanisms}
\bibinfo{author}{Bailey, D.W.}, \bibinfo{author}{Gross, J.E.},
  \bibinfo{author}{Laca, E.A.}, \bibinfo{author}{Rittenhouse, L.R.},
  \bibinfo{author}{Coughenour, M.B.}, \bibinfo{author}{Swift, D.M.},
  \bibinfo{author}{Sims, P.L.}, \bibinfo{year}{1996}.
\newblock \bibinfo{title}{Mechanisms that result in large herbivore grazing
  distribution patterns.}
\newblock \bibinfo{journal}{Rangeland Ecology \& Management/Journal of Range
  Management Archives} \bibinfo{volume}{49}, \bibinfo{pages}{386--400}.
\bibitem[{Beauchemin(2018)}]{beauchemin2018invited}
\bibinfo{author}{Beauchemin, K.}, \bibinfo{year}{2018}.
\newblock \bibinfo{title}{Invited review: Current perspectives on eating and
  rumination activity in dairy cows}.
\newblock \bibinfo{journal}{Journal of dairy science} \bibinfo{volume}{101},
  \bibinfo{pages}{4762--4784}.
\bibitem[{Beauchemin(1991)}]{Beauchemin1991-rp}
\bibinfo{author}{Beauchemin, K.A.}, \bibinfo{year}{1991}.
\newblock \bibinfo{title}{Ingestion and mastication of feed by dairy cattle}.
\newblock \bibinfo{journal}{Vet. Clin. North Am. Food Anim. Pract.}
  \bibinfo{volume}{7}, \bibinfo{pages}{439--463}.
\bibitem[{Benaissa et~al.(2019a)Benaissa, Tuyttens, Plets, Cattrysse, Martens,
  Vandaele, Joseph and Sonck}]{benaissa2019classification}
\bibinfo{author}{Benaissa, S.}, \bibinfo{author}{Tuyttens, F.A.},
  \bibinfo{author}{Plets, D.}, \bibinfo{author}{Cattrysse, H.},
  \bibinfo{author}{Martens, L.}, \bibinfo{author}{Vandaele, L.},
  \bibinfo{author}{Joseph, W.}, \bibinfo{author}{Sonck, B.},
  \bibinfo{year}{2019}a.
\newblock \bibinfo{title}{Classification of ingestive-related cow behaviours
  using rumiwatch halter and neck-mounted accelerometers}.
\newblock \bibinfo{journal}{Applied animal behaviour science}
  \bibinfo{volume}{211}, \bibinfo{pages}{9--16}.
\bibitem[{Benaissa et~al.(2019b)Benaissa, Tuyttens, Plets, De~Pessemier, Trogh,
  Tanghe, Martens, Vandaele, Van~Nuffel, Joseph et~al.}]{benaissa2019use}
\bibinfo{author}{Benaissa, S.}, \bibinfo{author}{Tuyttens, F.A.},
  \bibinfo{author}{Plets, D.}, \bibinfo{author}{De~Pessemier, T.},
  \bibinfo{author}{Trogh, J.}, \bibinfo{author}{Tanghe, E.},
  \bibinfo{author}{Martens, L.}, \bibinfo{author}{Vandaele, L.},
  \bibinfo{author}{Van~Nuffel, A.}, \bibinfo{author}{Joseph, W.}, et~al.,
  \bibinfo{year}{2019}b.
\newblock \bibinfo{title}{On the use of on-cow accelerometers for the
  classification of behaviours in dairy barns}.
\newblock \bibinfo{journal}{Research in veterinary science}
  \bibinfo{volume}{125}, \bibinfo{pages}{425--433}.
\bibitem[{Berckmans(2014)}]{Berckmans2014-qk}
\bibinfo{author}{Berckmans, D.}, \bibinfo{year}{2014}.
\newblock \bibinfo{title}{Precision livestock farming technologies for welfare
  management in intensive livestock systems}.
\newblock \bibinfo{journal}{Rev. Sci. Tech.} \bibinfo{volume}{33},
  \bibinfo{pages}{189--196}.
\bibitem[{Bishop(2006)}]{Bishop2006-ui}
\bibinfo{author}{Bishop, C.M.}, \bibinfo{year}{2006}.
\newblock \bibinfo{title}{Pattern Recognition and Machine Learning}.
\newblock \bibinfo{publisher}{Springer Verlag}.
\bibitem[{Bishop-Hurley et~al.(2014)Bishop-Hurley, Henry, Smith, Dutta, Hills,
  Rawnsley, Hellicar, Timms, Morshed, Rahman et~al.}]{bishop2014investigation}
\bibinfo{author}{Bishop-Hurley, G.}, \bibinfo{author}{Henry, D.},
  \bibinfo{author}{Smith, D.}, \bibinfo{author}{Dutta, R.},
  \bibinfo{author}{Hills, J.}, \bibinfo{author}{Rawnsley, R.},
  \bibinfo{author}{Hellicar, A.}, \bibinfo{author}{Timms, G.},
  \bibinfo{author}{Morshed, A.}, \bibinfo{author}{Rahman, A.}, et~al.,
  \bibinfo{year}{2014}.
\newblock \bibinfo{title}{An investigation of cow feeding behavior using motion
  sensors}, in: \bibinfo{booktitle}{2014 IEEE International Instrumentation and
  Measurement Technology Conference (I2MTC) Proceedings},
  \bibinfo{organization}{Ieee}. pp. \bibinfo{pages}{1285--1290}.
\bibitem[{Blair et~al.(1979)}]{blair1979information}
\bibinfo{author}{Blair, D.C.}, et~al., \bibinfo{year}{1979}.
\newblock \bibinfo{title}{Information retrieval, cj van rijsbergen. london:
  Butterworths; 1979: 208 pp.}
\newblock \bibinfo{journal}{Journal of the American Society for Information
  Science} \bibinfo{volume}{30}, \bibinfo{pages}{374--375}.
\bibitem[{Brennan et~al.(2021)Brennan, Johnson and
  Olson}]{brennan2021classifying}
\bibinfo{author}{Brennan, J.}, \bibinfo{author}{Johnson, P.},
  \bibinfo{author}{Olson, K.}, \bibinfo{year}{2021}.
\newblock \bibinfo{title}{Classifying season long livestock grazing behavior
  with the use of a low-cost gps and accelerometer}.
\newblock \bibinfo{journal}{Computers and Electronics in Agriculture}
  \bibinfo{volume}{181}, \bibinfo{pages}{105957}.
\bibitem[{Bristow and Holmes(2007)}]{Bristow2007-qe}
\bibinfo{author}{Bristow, D.J.}, \bibinfo{author}{Holmes, D.S.},
  \bibinfo{year}{2007}.
\newblock \bibinfo{title}{Cortisol levels and anxiety-related behaviors in
  cattle}.
\newblock \bibinfo{journal}{Physiol. Behav.} \bibinfo{volume}{90},
  \bibinfo{pages}{626--628}.
\bibitem[{Carslake et~al.(2020)Carslake, V{\'a}zquez-Diosdado and
  Kaler}]{carslake2020machine}
\bibinfo{author}{Carslake, C.}, \bibinfo{author}{V{\'a}zquez-Diosdado, J.A.},
  \bibinfo{author}{Kaler, J.}, \bibinfo{year}{2020}.
\newblock \bibinfo{title}{Machine learning algorithms to classify and quantify
  multiple behaviours in dairy calves using a sensor: Moving beyond
  classification in precision livestock}.
\newblock \bibinfo{journal}{Sensors} \bibinfo{volume}{21}, \bibinfo{pages}{88}.
\bibitem[{Chelotti et~al.(2018)Chelotti, Vanrell, Galli, Giovanini and
  Rufiner}]{chelotti2018pattern}
\bibinfo{author}{Chelotti, J.O.}, \bibinfo{author}{Vanrell, S.R.},
  \bibinfo{author}{Galli, J.R.}, \bibinfo{author}{Giovanini, L.L.},
  \bibinfo{author}{Rufiner, H.L.}, \bibinfo{year}{2018}.
\newblock \bibinfo{title}{A pattern recognition approach for detecting and
  classifying jaw movements in grazing cattle}.
\newblock \bibinfo{journal}{Computers and Electronics in Agriculture}
  \bibinfo{volume}{145}, \bibinfo{pages}{83--91}.
\bibitem[{Chelotti et~al.(2020)Chelotti, Vanrell, Rau, Galli, Planisich,
  Utsumi, Milone, Giovanini and Rufiner}]{chelotti2020online}
\bibinfo{author}{Chelotti, J.O.}, \bibinfo{author}{Vanrell, S.R.},
  \bibinfo{author}{Rau, L.S.M.}, \bibinfo{author}{Galli, J.R.},
  \bibinfo{author}{Planisich, A.M.}, \bibinfo{author}{Utsumi, S.A.},
  \bibinfo{author}{Milone, D.H.}, \bibinfo{author}{Giovanini, L.L.},
  \bibinfo{author}{Rufiner, H.L.}, \bibinfo{year}{2020}.
\newblock \bibinfo{title}{An online method for estimating grazing and
  rumination bouts using acoustic signals in grazing cattle}.
\newblock \bibinfo{journal}{Computers and Electronics in Agriculture}
  \bibinfo{volume}{173}, \bibinfo{pages}{105443}.
\bibitem[{Chen et~al.(2022)Chen, Li, Guo, Shu, Cao and
  Xu}]{chen2022recognition}
\bibinfo{author}{Chen, G.}, \bibinfo{author}{Li, C.}, \bibinfo{author}{Guo,
  Y.}, \bibinfo{author}{Shu, H.}, \bibinfo{author}{Cao, Z.},
  \bibinfo{author}{Xu, B.}, \bibinfo{year}{2022}.
\newblock \bibinfo{title}{Recognition of cattle's feeding behaviors using
  noseband pressure sensor with machine learning}.
\newblock \bibinfo{journal}{Frontiers in Veterinary Science}
  \bibinfo{volume}{9}.
\bibitem[{Dutta et~al.(2015)Dutta, Smith, Rawnsley, Bishop-Hurley, Hills, Timms
  and Henry}]{dutta2015dynamic}
\bibinfo{author}{Dutta, R.}, \bibinfo{author}{Smith, D.},
  \bibinfo{author}{Rawnsley, R.}, \bibinfo{author}{Bishop-Hurley, G.},
  \bibinfo{author}{Hills, J.}, \bibinfo{author}{Timms, G.},
  \bibinfo{author}{Henry, D.}, \bibinfo{year}{2015}.
\newblock \bibinfo{title}{Dynamic cattle behavioural classification using
  supervised ensemble classifiers}.
\newblock \bibinfo{journal}{Computers and electronics in agriculture}
  \bibinfo{volume}{111}, \bibinfo{pages}{18--28}.
\bibitem[{Eslamizad et~al.(2018)Eslamizad, T{\"u}mmler, Derno, Hoch and
  Kuhla}]{eslamizad2018development}
\bibinfo{author}{Eslamizad, M.}, \bibinfo{author}{T{\"u}mmler, L.M.},
  \bibinfo{author}{Derno, M.}, \bibinfo{author}{Hoch, M.},
  \bibinfo{author}{Kuhla, B.}, \bibinfo{year}{2018}.
\newblock \bibinfo{title}{Development of a pressure sensor-based system for
  measuring rumination time in pre-weaned dairy calves}.
\newblock \bibinfo{journal}{Journal of Animal Science} \bibinfo{volume}{96},
  \bibinfo{pages}{4483--4489}.
\bibitem[{Farr{\'u}s et~al.(2007)Farr{\'u}s, Hernando and
  Ejarque}]{farrus2007jitter}
\bibinfo{author}{Farr{\'u}s, M.}, \bibinfo{author}{Hernando, J.},
  \bibinfo{author}{Ejarque, P.}, \bibinfo{year}{2007}.
\newblock \bibinfo{title}{Jitter and shimmer measurements for speaker
  recognition}, in: \bibinfo{booktitle}{8th Annual Conference of the
  International Speech Communication Association; 2007 Aug. 27-31; Antwerp
  (Belgium).[place unknown]: ISCA; 2007. p. 778-81.},
  \bibinfo{organization}{International Speech Communication Association
  (ISCA)}.
\bibitem[{Galli(2008)}]{galli2008Thesis}
\bibinfo{author}{Galli, J.R.}, \bibinfo{year}{2008}.
\newblock \bibinfo{title}{Medición acústica del comportamiento ingestivo y
  del consumo de rumiantes en pastoreo.}
\newblock \bibinfo{journal}{Tesis Doctorado en Ciencias Agrarias –
  Orientación Producción Animal. Facultad de Ciencias Agrarias, Universidad
  Nacional de Mar del Plata, Balcarce, Argentina} .
\bibitem[{Galli et~al.(2011)Galli, Cangiano, Milone and
  Laca}]{galli2011acoustic}
\bibinfo{author}{Galli, J.R.}, \bibinfo{author}{Cangiano, C.A.},
  \bibinfo{author}{Milone, D.H.}, \bibinfo{author}{Laca, E.A.},
  \bibinfo{year}{2011}.
\newblock \bibinfo{title}{Acoustic monitoring of short-term ingestive behavior
  and intake in grazing sheep}.
\newblock \bibinfo{journal}{Livestock Science} \bibinfo{volume}{140},
  \bibinfo{pages}{32--41}.
\bibitem[{Galli et~al.(2018)Galli, Cangiano, Pece, Larripa, Milone, Utsumi and
  Laca}]{Galli2018-zq}
\bibinfo{author}{Galli, J.R.}, \bibinfo{author}{Cangiano, C.A.},
  \bibinfo{author}{Pece, M.A.}, \bibinfo{author}{Larripa, M.J.},
  \bibinfo{author}{Milone, D.H.}, \bibinfo{author}{Utsumi, S.A.},
  \bibinfo{author}{Laca, E.A.}, \bibinfo{year}{2018}.
\newblock \bibinfo{title}{Monitoring and assessment of ingestive chewing sounds
  for prediction of herbage intake rate in grazing cattle}.
\newblock \bibinfo{journal}{Animal} \bibinfo{volume}{12},
  \bibinfo{pages}{973--982}.
\bibitem[{Galli et~al.(2020)Galli, Milone, Cangiano, Mart{\'\i}nez, Laca,
  Chelotti and Rufiner}]{galli2020discriminative}
\bibinfo{author}{Galli, J.R.}, \bibinfo{author}{Milone, D.H.},
  \bibinfo{author}{Cangiano, C.A.}, \bibinfo{author}{Mart{\'\i}nez, C.E.},
  \bibinfo{author}{Laca, E.A.}, \bibinfo{author}{Chelotti, J.O.},
  \bibinfo{author}{Rufiner, H.L.}, \bibinfo{year}{2020}.
\newblock \bibinfo{title}{Discriminative power of acoustic features for jaw
  movement classification in cattle and sheep}.
\newblock \bibinfo{journal}{Bioacoustics} \bibinfo{volume}{29},
  \bibinfo{pages}{602--616}.
\bibitem[{Herskin et~al.(2004)Herskin, Munksgaard and Ladewig}]{Herskin2004-km}
\bibinfo{author}{Herskin, M.S.}, \bibinfo{author}{Munksgaard, L.},
  \bibinfo{author}{Ladewig, J.}, \bibinfo{year}{2004}.
\newblock \bibinfo{title}{Effects of acute stressors on nociception,
  adrenocortical responses and behavior of dairy cows}.
\newblock \bibinfo{journal}{Physiol. Behav.} \bibinfo{volume}{83},
  \bibinfo{pages}{411--420}.
\bibitem[{Hodgson and Illius(1998)}]{Hodgson1998-xy}
\bibinfo{author}{Hodgson, J.G.}, \bibinfo{author}{Illius, A.W.},
  \bibinfo{year}{1998}.
\newblock \bibinfo{title}{The Ecology and Management of Grazing Systems}.
\newblock \bibinfo{publisher}{C A B International}.
\bibitem[{Kamminga et~al.(2018)Kamminga, Le, Meijers, Bisby, Meratnia and
  Havinga}]{kamminga2018robust}
\bibinfo{author}{Kamminga, J.W.}, \bibinfo{author}{Le, D.V.},
  \bibinfo{author}{Meijers, J.P.}, \bibinfo{author}{Bisby, H.},
  \bibinfo{author}{Meratnia, N.}, \bibinfo{author}{Havinga, P.J.},
  \bibinfo{year}{2018}.
\newblock \bibinfo{title}{Robust sensor-orientation-independent feature
  selection for animal activity recognition on collar tags}.
\newblock \bibinfo{journal}{Proceedings of the ACM on Interactive, Mobile,
  Wearable and Ubiquitous Technologies} \bibinfo{volume}{2},
  \bibinfo{pages}{1--27}.
\bibitem[{Kohavi and John(1997)}]{kohavi1997wrappers}
\bibinfo{author}{Kohavi, R.}, \bibinfo{author}{John, G.H.},
  \bibinfo{year}{1997}.
\newblock \bibinfo{title}{Wrappers for feature subset selection}.
\newblock \bibinfo{journal}{Artificial intelligence} \bibinfo{volume}{97},
  \bibinfo{pages}{273--324}.
\bibitem[{Laca et~al.(1992)Laca, Ungar, Seligman, Ramey and
  Demment}]{Laca1992-qq}
\bibinfo{author}{Laca, E.A.}, \bibinfo{author}{Ungar, E.D.},
  \bibinfo{author}{Seligman, N.G.}, \bibinfo{author}{Ramey, M.R.},
  \bibinfo{author}{Demment, M.W.}, \bibinfo{year}{1992}.
\newblock \bibinfo{title}{An integrated methodology for studying short-term
  grazing behaviour of cattle}.
\newblock \bibinfo{journal}{Grass Forage Sci.} \bibinfo{volume}{47},
  \bibinfo{pages}{81--90}.
\bibitem[{Laca et~al.(2000)Laca, WallisDeVries et~al.}]{laca2000acoustic}
\bibinfo{author}{Laca, E.A.}, \bibinfo{author}{WallisDeVries, M.F.}, et~al.,
  \bibinfo{year}{2000}.
\newblock \bibinfo{title}{Acoustic measurement of intake and grazing behaviour
  of cattle.}
\newblock \bibinfo{journal}{Grass and Forage Science} \bibinfo{volume}{55},
  \bibinfo{pages}{97--104}.
\bibitem[{Lee and Seo(2021)}]{lee2021wearable}
\bibinfo{author}{Lee, M.}, \bibinfo{author}{Seo, S.}, \bibinfo{year}{2021}.
\newblock \bibinfo{title}{Wearable wireless biosensor technology for monitoring
  cattle: A review}.
\newblock \bibinfo{journal}{Animals} \bibinfo{volume}{11},
  \bibinfo{pages}{2779}.
\bibitem[{Li et~al.(2021a)Li, Tokgoz, Fukawa, Bartels, Ohashi, Takeda and
  Ito}]{li2021data}
\bibinfo{author}{Li, C.}, \bibinfo{author}{Tokgoz, K.K.},
  \bibinfo{author}{Fukawa, M.}, \bibinfo{author}{Bartels, J.},
  \bibinfo{author}{Ohashi, T.}, \bibinfo{author}{Takeda, K.i.},
  \bibinfo{author}{Ito, H.}, \bibinfo{year}{2021}a.
\newblock \bibinfo{title}{Data augmentation for inertial sensor data in cnns
  for cattle behavior classification}.
\newblock \bibinfo{journal}{IEEE Sensors Letters} \bibinfo{volume}{5},
  \bibinfo{pages}{1--4}.
\bibitem[{Li et~al.(2021b)Li, Xiong, Du, Shi and Gates}]{li2021}
\bibinfo{author}{Li, G.}, \bibinfo{author}{Xiong, Y.}, \bibinfo{author}{Du,
  Q.}, \bibinfo{author}{Shi, Z.}, \bibinfo{author}{Gates, R.S.},
  \bibinfo{year}{2021}b.
\newblock \bibinfo{title}{Classifying ingestive behavior of dairy cows via
  automatic sound recognition}.
\newblock \bibinfo{journal}{Sensors} \bibinfo{volume}{21}.
\newblock \URLprefix \url{https://www.mdpi.com/1424-8220/21/15/5231},
  \DOIprefix\doi{10.3390/s21155231}.
\bibitem[{Li et~al.(2022)Li, Shu, Bindelle, Xu, Zhang, Jin, Guo and
  Wang}]{li2022classification}
\bibinfo{author}{Li, Y.}, \bibinfo{author}{Shu, H.}, \bibinfo{author}{Bindelle,
  J.}, \bibinfo{author}{Xu, B.}, \bibinfo{author}{Zhang, W.},
  \bibinfo{author}{Jin, Z.}, \bibinfo{author}{Guo, L.}, \bibinfo{author}{Wang,
  W.}, \bibinfo{year}{2022}.
\newblock \bibinfo{title}{Classification and analysis of multiple cattle
  unitary behaviors and movements based on machine learning methods}.
\newblock \bibinfo{journal}{Animals} \bibinfo{volume}{12},
  \bibinfo{pages}{1060}.
\bibitem[{Van~der Maaten and Hinton(2012)}]{van2012visualizing}
\bibinfo{author}{Van~der Maaten, L.}, \bibinfo{author}{Hinton, G.},
  \bibinfo{year}{2012}.
\newblock \bibinfo{title}{Visualizing non-metric similarities in multiple
  maps}.
\newblock \bibinfo{journal}{Machine learning} \bibinfo{volume}{87},
  \bibinfo{pages}{33--55}.
\bibitem[{Martinez-Rau et~al.(2022)Martinez-Rau, Chelotti, Vanrell, Galli,
  Utsumi, Planisich, Rufiner and Giovanini}]{martinez2022robust}
\bibinfo{author}{Martinez-Rau, L.S.}, \bibinfo{author}{Chelotti, J.O.},
  \bibinfo{author}{Vanrell, S.R.}, \bibinfo{author}{Galli, J.R.},
  \bibinfo{author}{Utsumi, S.A.}, \bibinfo{author}{Planisich, A.M.},
  \bibinfo{author}{Rufiner, H.L.}, \bibinfo{author}{Giovanini, L.L.},
  \bibinfo{year}{2022}.
\newblock \bibinfo{title}{A robust computational approach for jaw movement
  detection and classification in grazing cattle using acoustic signals}.
\newblock \bibinfo{journal}{Computers and Electronics in Agriculture}
  \bibinfo{volume}{192}, \bibinfo{pages}{106569}.
\bibitem[{Pavlovic et~al.(2022)Pavlovic, Czerkawski, Davison, Marko, Michie,
  Atkinson, Crnojevic, Andonovic, Rajovic, Kvascev
  et~al.}]{pavlovic2022behavioural}
\bibinfo{author}{Pavlovic, D.}, \bibinfo{author}{Czerkawski, M.},
  \bibinfo{author}{Davison, C.}, \bibinfo{author}{Marko, O.},
  \bibinfo{author}{Michie, C.}, \bibinfo{author}{Atkinson, R.},
  \bibinfo{author}{Crnojevic, V.}, \bibinfo{author}{Andonovic, I.},
  \bibinfo{author}{Rajovic, V.}, \bibinfo{author}{Kvascev, G.}, et~al.,
  \bibinfo{year}{2022}.
\newblock \bibinfo{title}{Behavioural classification of cattle using
  neck-mounted accelerometer-equipped collars}.
\newblock \bibinfo{journal}{Sensors} \bibinfo{volume}{22},
  \bibinfo{pages}{2323}.
\bibitem[{Pavlovic et~al.(2021)Pavlovic, Davison, Hamilton, Marko, Atkinson,
  Michie, Crnojevi{\'c}, Andonovic, Bellekens and
  Tachtatzis}]{pavlovic2021classification}
\bibinfo{author}{Pavlovic, D.}, \bibinfo{author}{Davison, C.},
  \bibinfo{author}{Hamilton, A.}, \bibinfo{author}{Marko, O.},
  \bibinfo{author}{Atkinson, R.}, \bibinfo{author}{Michie, C.},
  \bibinfo{author}{Crnojevi{\'c}, V.}, \bibinfo{author}{Andonovic, I.},
  \bibinfo{author}{Bellekens, X.}, \bibinfo{author}{Tachtatzis, C.},
  \bibinfo{year}{2021}.
\newblock \bibinfo{title}{Classification of cattle behaviours using
  neck-mounted accelerometer-equipped collars and convolutional neural
  networks}.
\newblock \bibinfo{journal}{Sensors} \bibinfo{volume}{21},
  \bibinfo{pages}{4050}.
\bibitem[{Peng et~al.(2019)Peng, Kondo, Fujiura, Suzuki, Yoshioka, Itoyama
  et~al.}]{peng2019classification}
\bibinfo{author}{Peng, Y.}, \bibinfo{author}{Kondo, N.},
  \bibinfo{author}{Fujiura, T.}, \bibinfo{author}{Suzuki, T.},
  \bibinfo{author}{Yoshioka, H.}, \bibinfo{author}{Itoyama, E.}, et~al.,
  \bibinfo{year}{2019}.
\newblock \bibinfo{title}{Classification of multiple cattle behavior patterns
  using a recurrent neural network with long short-term memory and inertial
  measurement units}.
\newblock \bibinfo{journal}{Computers and electronics in agriculture}
  \bibinfo{volume}{157}, \bibinfo{pages}{247--253}.
\bibitem[{Rahman et~al.(2016)Rahman, Smith, Hills, Bishop-Hurley, Henry and
  Rawnsley}]{rahman2016comparison}
\bibinfo{author}{Rahman, A.}, \bibinfo{author}{Smith, D.},
  \bibinfo{author}{Hills, J.}, \bibinfo{author}{Bishop-Hurley, G.},
  \bibinfo{author}{Henry, D.}, \bibinfo{author}{Rawnsley, R.},
  \bibinfo{year}{2016}.
\newblock \bibinfo{title}{A comparison of autoencoder and statistical features
  for cattle behaviour classification}, in: \bibinfo{booktitle}{2016
  international joint conference on neural networks (IJCNN)},
  \bibinfo{organization}{IEEE}. pp. \bibinfo{pages}{2954--2960}.
\bibitem[{Rahman et~al.(2018)Rahman, Smith, Little, Ingham, Greenwood and
  Bishop-Hurley}]{rahman2018cattle}
\bibinfo{author}{Rahman, A.}, \bibinfo{author}{Smith, D.},
  \bibinfo{author}{Little, B.}, \bibinfo{author}{Ingham, A.},
  \bibinfo{author}{Greenwood, P.}, \bibinfo{author}{Bishop-Hurley, G.},
  \bibinfo{year}{2018}.
\newblock \bibinfo{title}{Cattle behaviour classification from collar, halter,
  and ear tag sensors}.
\newblock \bibinfo{journal}{Information processing in agriculture}
  \bibinfo{volume}{5}, \bibinfo{pages}{124--133}.
\bibitem[{Riaboff et~al.(2020)Riaboff, Poggi, Madouasse, Couvreur, Aubin,
  B{\'e}d{\`e}re, Goumand, Chauvin and Plantier}]{riaboff2020development}
\bibinfo{author}{Riaboff, L.}, \bibinfo{author}{Poggi, S.},
  \bibinfo{author}{Madouasse, A.}, \bibinfo{author}{Couvreur, S.},
  \bibinfo{author}{Aubin, S.}, \bibinfo{author}{B{\'e}d{\`e}re, N.},
  \bibinfo{author}{Goumand, E.}, \bibinfo{author}{Chauvin, A.},
  \bibinfo{author}{Plantier, G.}, \bibinfo{year}{2020}.
\newblock \bibinfo{title}{Development of a methodological framework for a
  robust prediction of the main behaviours of dairy cows using a combination of
  machine learning algorithms on accelerometer data}.
\newblock \bibinfo{journal}{Computers and Electronics in Agriculture}
  \bibinfo{volume}{169}, \bibinfo{pages}{105179}.
\bibitem[{Riaboff et~al.(2022)Riaboff, Shalloo, Smeaton, Couvreur, Madouasse
  and Keane}]{riaboff2022predicting}
\bibinfo{author}{Riaboff, L.}, \bibinfo{author}{Shalloo, L.},
  \bibinfo{author}{Smeaton, A.F.}, \bibinfo{author}{Couvreur, S.},
  \bibinfo{author}{Madouasse, A.}, \bibinfo{author}{Keane, M.T.},
  \bibinfo{year}{2022}.
\newblock \bibinfo{title}{Predicting livestock behaviour using accelerometers:
  A systematic review of processing techniques for ruminant behaviour
  prediction from raw accelerometer data}.
\newblock \bibinfo{journal}{Computers and Electronics in Agriculture}
  \bibinfo{volume}{192}, \bibinfo{pages}{106610}.
\bibitem[{Sakai et~al.(2019)Sakai, Oishi, Miwa, Kumagai and
  Hirooka}]{sakai2019behavior}
\bibinfo{author}{Sakai, K.}, \bibinfo{author}{Oishi, K.},
  \bibinfo{author}{Miwa, M.}, \bibinfo{author}{Kumagai, H.},
  \bibinfo{author}{Hirooka, H.}, \bibinfo{year}{2019}.
\newblock \bibinfo{title}{Behavior classification of goats using 9-axis multi
  sensors: The effect of imbalanced datasets on classification performance}.
\newblock \bibinfo{journal}{Computers and Electronics in Agriculture}
  \bibinfo{volume}{166}, \bibinfo{pages}{105027}.
\bibitem[{Schirmann et~al.(2013)Schirmann, Chapinal, Weary, Vickers and von
  Keyserlingk}]{Schirmann2013-yq}
\bibinfo{author}{Schirmann, K.}, \bibinfo{author}{Chapinal, N.},
  \bibinfo{author}{Weary, D.M.}, \bibinfo{author}{Vickers, L.},
  \bibinfo{author}{von Keyserlingk, M.A.G.}, \bibinfo{year}{2013}.
\newblock \bibinfo{title}{Short communication: Rumination and feeding behavior
  before and after calving in dairy cows}.
\newblock \bibinfo{journal}{J. Dairy Sci.} \bibinfo{volume}{96},
  \bibinfo{pages}{7088--7092}.
\bibitem[{Schirmann et~al.(2009)Schirmann, von Keyserlingk, Weary, Veira and
  Heuwieser}]{Schirmann2009-ns}
\bibinfo{author}{Schirmann, K.}, \bibinfo{author}{von Keyserlingk, M.A.G.},
  \bibinfo{author}{Weary, D.M.}, \bibinfo{author}{Veira, D.M.},
  \bibinfo{author}{Heuwieser, W.}, \bibinfo{year}{2009}.
\newblock \bibinfo{title}{Technical note: Validation of a system for monitoring
  rumination in dairy cows}.
\newblock \bibinfo{journal}{J. Dairy Sci.} \bibinfo{volume}{92},
  \bibinfo{pages}{6052--6055}.
\bibitem[{Shen et~al.(2020)Shen, Cheng, Zhang, Wei, Fu and
  Zhang}]{shen2020automatic}
\bibinfo{author}{Shen, W.}, \bibinfo{author}{Cheng, F.},
  \bibinfo{author}{Zhang, Y.}, \bibinfo{author}{Wei, X.}, \bibinfo{author}{Fu,
  Q.}, \bibinfo{author}{Zhang, Y.}, \bibinfo{year}{2020}.
\newblock \bibinfo{title}{Automatic recognition of ingestive-related behaviors
  of dairy cows based on triaxial acceleration}.
\newblock \bibinfo{journal}{Information Processing in Agriculture}
  \bibinfo{volume}{7}, \bibinfo{pages}{427--443}.
\bibitem[{Smith et~al.(2016)Smith, Rahman, Bishop-Hurley, Hills, Shahriar,
  Henry and Rawnsley}]{Smith2016-xb}
\bibinfo{author}{Smith, D.}, \bibinfo{author}{Rahman, A.},
  \bibinfo{author}{Bishop-Hurley, G.J.}, \bibinfo{author}{Hills, J.},
  \bibinfo{author}{Shahriar, S.}, \bibinfo{author}{Henry, D.},
  \bibinfo{author}{Rawnsley, R.}, \bibinfo{year}{2016}.
\newblock \bibinfo{title}{Behavior classification of cows fitted with motion
  collars: Decomposing multi-class classification into a set of binary
  problems}.
\newblock \bibinfo{journal}{Comput. Electron. Agric.} \bibinfo{volume}{131},
  \bibinfo{pages}{40--50}.
\bibitem[{Steinmetz et~al.(2020)Steinmetz, von Soosten, Hummel, Meyer and
  D{\"a}nicke}]{steinmetz2020validation}
\bibinfo{author}{Steinmetz, M.}, \bibinfo{author}{von Soosten, D.},
  \bibinfo{author}{Hummel, J.}, \bibinfo{author}{Meyer, U.},
  \bibinfo{author}{D{\"a}nicke, S.}, \bibinfo{year}{2020}.
\newblock \bibinfo{title}{Validation of the rumiwatch converter v0. 7.4. 5
  classification accuracy for the automatic monitoring of behavioural
  characteristics in dairy cows}.
\newblock \bibinfo{journal}{Archives of animal nutrition} \bibinfo{volume}{74},
  \bibinfo{pages}{164--172}.
\bibitem[{Stone(2020)}]{Stone2020-jl}
\bibinfo{author}{Stone, A.E.}, \bibinfo{year}{2020}.
\newblock \bibinfo{title}{Symposium review: The most important factors
  affecting adoption of precision dairy monitoring technologies}.
\newblock \bibinfo{journal}{J. Dairy Sci.} \bibinfo{volume}{103},
  \bibinfo{pages}{5740--5745}.
\bibitem[{Ungar and Rutter(2006)}]{Ungar2006-ei}
\bibinfo{author}{Ungar, E.D.}, \bibinfo{author}{Rutter, S.M.},
  \bibinfo{year}{2006}.
\newblock \bibinfo{title}{Classifying cattle jaw movements: Comparing {IGER}
  behaviour recorder and acoustic techniques}.
\newblock \bibinfo{journal}{Appl. Anim. Behav. Sci.} \bibinfo{volume}{98},
  \bibinfo{pages}{11--27}.
\bibitem[{Vanrell et~al.(2018)Vanrell, Chelotti, Galli, Utsumi, Giovanini,
  Rufiner and Milone}]{Vanrell2018-zj}
\bibinfo{author}{Vanrell, S.R.}, \bibinfo{author}{Chelotti, J.O.},
  \bibinfo{author}{Galli, J.R.}, \bibinfo{author}{Utsumi, S.A.},
  \bibinfo{author}{Giovanini, L.L.}, \bibinfo{author}{Rufiner, H.L.},
  \bibinfo{author}{Milone, D.H.}, \bibinfo{year}{2018}.
\newblock \bibinfo{title}{A regularity-based algorithm for identifying grazing
  and rumination bouts from acoustic signals in grazing cattle}.
\newblock \bibinfo{journal}{Comput. Electron. Agric.} \bibinfo{volume}{151},
  \bibinfo{pages}{392--402}.
\bibitem[{Welch(1982)}]{welch1982rumination}
\bibinfo{author}{Welch, J.}, \bibinfo{year}{1982}.
\newblock \bibinfo{title}{Rumination, particle size and passage from the
  rumen}.
\newblock \bibinfo{journal}{Journal of animal science} \bibinfo{volume}{54},
  \bibinfo{pages}{885--894}.
\bibitem[{Werner et~al.(2018)Werner, Leso, Umstatter, Niederhauser, Kennedy,
  Geoghegan, Shalloo, Schick and O’Brien}]{werner2018evaluation}
\bibinfo{author}{Werner, J.}, \bibinfo{author}{Leso, L.},
  \bibinfo{author}{Umstatter, C.}, \bibinfo{author}{Niederhauser, J.},
  \bibinfo{author}{Kennedy, E.}, \bibinfo{author}{Geoghegan, A.},
  \bibinfo{author}{Shalloo, L.}, \bibinfo{author}{Schick, M.},
  \bibinfo{author}{O’Brien, B.}, \bibinfo{year}{2018}.
\newblock \bibinfo{title}{Evaluation of the
  \uppercase{R}umi\uppercase{W}atch\uppercase{S}ystem for measuring grazing
  behaviour of cows}.
\newblock \bibinfo{journal}{Journal of Neuroscience Methods}
  \bibinfo{volume}{300}, \bibinfo{pages}{138--146}.
\bibitem[{Widrow et~al.(1975)Widrow, Glover, McCool, Kaunitz, Williams, Hearn,
  Zeidler, Dong and Goodlin}]{widrow1975adaptive}
\bibinfo{author}{Widrow, B.}, \bibinfo{author}{Glover, J.R.},
  \bibinfo{author}{McCool, J.M.}, \bibinfo{author}{Kaunitz, J.},
  \bibinfo{author}{Williams, C.S.}, \bibinfo{author}{Hearn, R.H.},
  \bibinfo{author}{Zeidler, J.R.}, \bibinfo{author}{Dong, J.E.},
  \bibinfo{author}{Goodlin, R.C.}, \bibinfo{year}{1975}.
\newblock \bibinfo{title}{Adaptive noise cancelling: Principles and
  applications}.
\newblock \bibinfo{journal}{Proceedings of the IEEE} \bibinfo{volume}{63},
  \bibinfo{pages}{1692--1716}.
\bibitem[{Wilcoxon(1947)}]{wilcoxon1947probability}
\bibinfo{author}{Wilcoxon, F.}, \bibinfo{year}{1947}.
\newblock \bibinfo{title}{Probability tables for individual comparisons by
  ranking methods}.
\newblock \bibinfo{journal}{Biometrics} \bibinfo{volume}{3},
  \bibinfo{pages}{119--122}.
\bibitem[{Zehner et~al.(2017)Zehner, Umst{\"a}tter, Niederhauser and
  Schick}]{zehner2017system}
\bibinfo{author}{Zehner, N.}, \bibinfo{author}{Umst{\"a}tter, C.},
  \bibinfo{author}{Niederhauser, J.J.}, \bibinfo{author}{Schick, M.},
  \bibinfo{year}{2017}.
\newblock \bibinfo{title}{System specification and validation of a noseband
  pressure sensor for measurement of ruminating and eating behavior in
  stable-fed cows}.
\newblock \bibinfo{journal}{Computers and Electronics in Agriculture}
  \bibinfo{volume}{136}, \bibinfo{pages}{31--41}.

\end{thebibliography}







\end{document}


\section*{Supplementary Material: Comparison of \mbox{JMFAR} and \mbox{BUFAR}}

\begin{figure}[h]
       \centering
       \includegraphics[width=.9\textwidth]{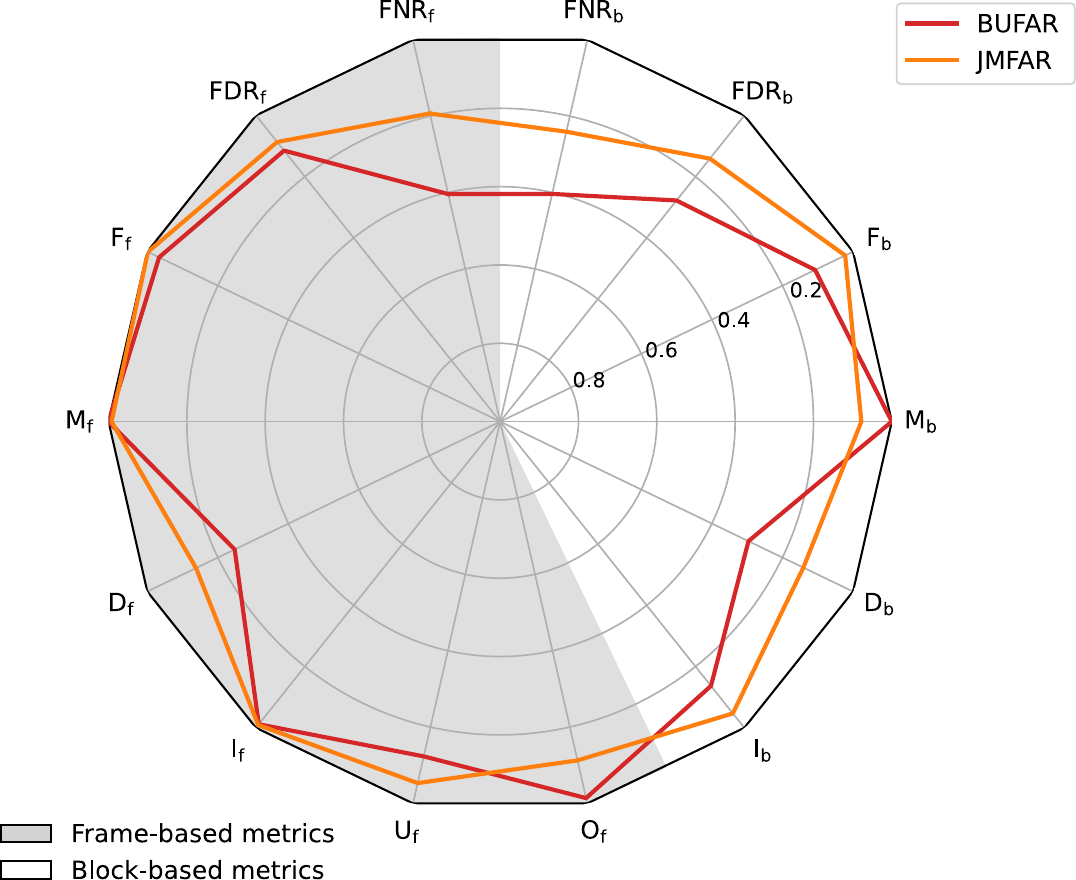}
       \caption{Spider plot of frame- and block-based metrics for grazing classification for the former (\mbox{BUFAR}) and the proposed method (\mbox{JMFAR}). Error metrics are: false negative rate (FNR), false discovery rate (FDR), fragmentation (F), merging (M), deletion (D), insertion (I), underfill (U) and overfill (O). The subscript indicates frame (f) or block-based (b) metrics.}
       \label{fig:grazing}
\end{figure}

\begin{figure}[h]
       \centering
       \includegraphics[width=.9\textwidth]	{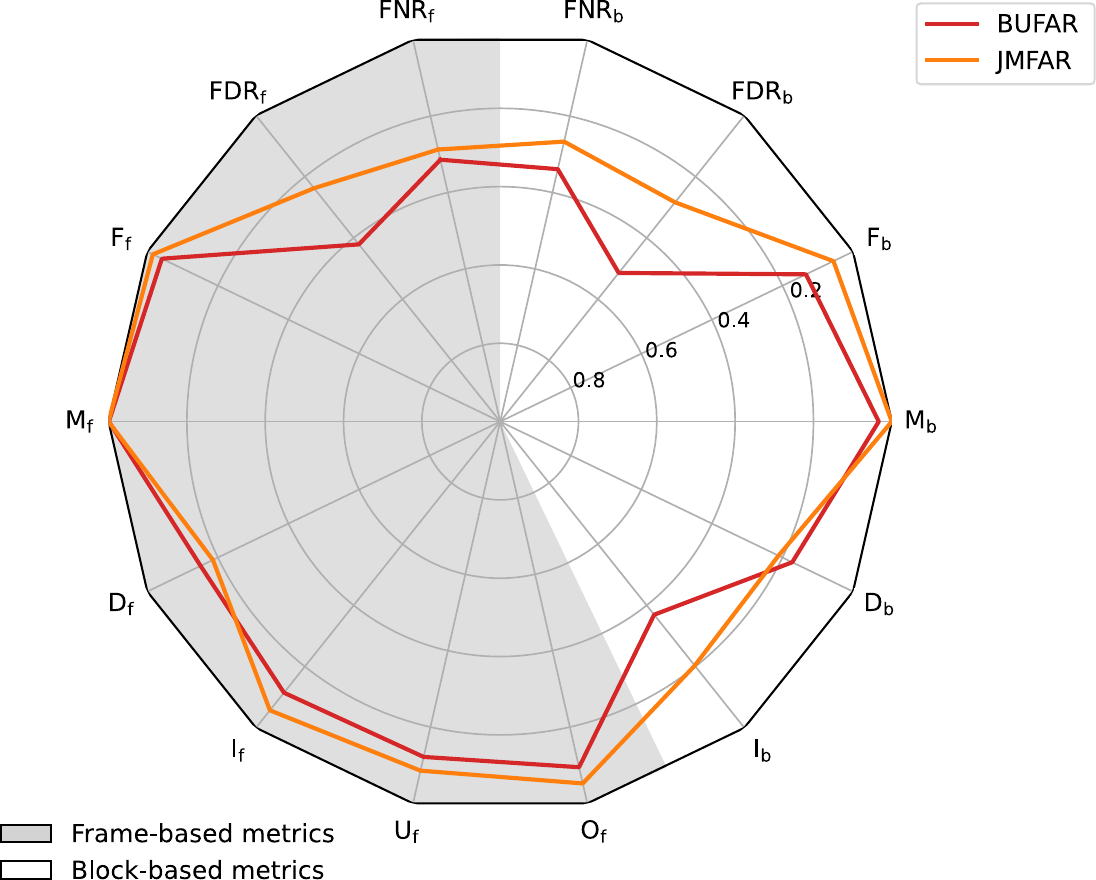}
       \caption{Spider plot of frame- and block-based metrics for rumination classification for the former method (\mbox{BUFAR}) and the proposed method (\mbox{JMFAR}). Error metrics are: false negative rate (FNR), false discovery rate (FDR), fragmentation (F), merging (M), deletion (D), insertion (I), underfill (U) and overfill (O). The subscript indicates frame (f) or block-based (b) metrics.}
       \label{fig:rumination}
\end{figure}

The spiders plots in Figures~\ref{fig:grazing} and~\ref{fig:rumination} summarize the performance for \mbox{JMFAR} and \mbox{BUFAR} with the metrics used and detailed in \citet{Vanrell2018-zj} and \citet{chelotti2020online}.

\subsection*{Grazing}

\subsection*{Rumination}

\bibliographystyle{elsarticle-harv}
\bibliography{references.bib}